\documentclass[11pt,a4paper]{article}
\usepackage{amsmath,amsfonts,amssymb,amsthm,amstext,amscd,array,bbold}
\DeclareMathOperator{\AdS}{AdS}
\DeclareMathOperator{\Sphere}{S}
\DeclareMathOperator{\NW}{NW}
\DeclareMathOperator{\CW}{CW}
\let\S\Sphere
\DeclareMathOperator{\weyl}{{\sf W}}                   

\newcommand{\comp}{\boxplus}                            
\newcommand{\compa}{\mkern+4mu\square\mkern-17.4mu{\mbox{
    \protect\raisebox{0.2ex}{$\ast$}}\mkern+5mu}}       
\newcommand{\compb}{\mkern+4mu\square\mkern-17.4mu{\mbox{
    \protect\raisebox{0.2ex}{$\bullet$}}\mkern+5mu}}    
\newcommand{\compc}{\mkern+4mu\square\mkern-17.4mu{\mbox{
    \protect\raisebox{0.2ex}{$\star$}}\mkern+5mu}}      
\newcommand{\cb}[2]{\bigl[#1\,,\,#2\bigr]}                 
\newcommand{\NO}{\mbox{$\substack{\circ\\\circ}$}}      
\newcommand{\NOa}{\mbox{$\substack{\ast\\\ast}$}}       
\newcommand{\NOb}{\mbox{$\substack{\bullet\\\bullet}$}} 

\def\nn{\nonumber}
\def\d{\partial}
\def\bfd{{\mbf\partial}}
\def\bfz{{\mbf z}}
\def\od{\overline{\partial}}
\def\oz{\overline{z}}
\def\obfd{\overline{\mbf\partial}}
\def\obfz{\overline{\mbf z}}

\newcommand{\N}{\nat}

\def\one{\mathbb{1}}

\newcommand{\mbf}[1]{{\boldsymbol {#1} }}
\def\ii{{\,{\rm i}\,}}
\def\dd{{\rm d}}

\def\CC{{\rm C}}
\def\P{{\sf P}}

\def\T{{\sf T}}
\def\X{{\sf X}}
\def\Y{{\sf Y}}

\def\J{{\sf J}}

\def\ma{{\mbf a}}
\def\mz{{\mbf z}}
\def\mw{{\mbf w}}
\def\mx{{\mbf x}}

\def\mk{{\mbf k}}
\def\mq{{\mbf q}}
\def\mbp{{\mbf p}}
\def\mD{{\mbf D}}
\def\mG{{\mbf G}}
\def\mdell{{\mbf\partial}}

\def\mfn{{\mathfrak n}}
\def\mfg{{\mathfrak g}}
\def\mfs{{\mathfrak s}}
\def\mcN{{\mathcal N}}

\def\mbbV{{\mathbb V}}

\newcommand{\ncint}{\int\!\!\!\!\!\! - ~}

\newcommand{\eq}{\begin{equation}}
\newcommand{\eqend}{\end{equation}}
\newcommand{\eqa}{\begin{eqnarray}}
\newcommand{\nonueqa}{\begin{eqnarray*}}
\newcommand{\eqaend}{\end{eqnarray}}
\newcommand{\nonueqaend}{\end{eqnarray*}}

\newcommand{\bma}[1]{\begin{array}{#1}}
\newcommand{\ema}{\end{array}}
\newcommand{\bc}{\begin{center}}
\newcommand{\ec}{\end{center}}

\newcommand{\R}{\real}

\renewcommand{\thefootnote}{\fnsymbol{footnote}}
\newcommand{\newsection}{\setcounter{equation}{0}\section}

\def\appendix#1{\addtocounter{section}{1}\setcounter{equation}{0}
\renewcommand{\thesection}{\Alph{section}}
\section*{Appendix \thesection\protect\indent \parbox[t]{11.715cm} {#1}}
\addcontentsline{toc}{section}{Appendix \thesection\ \ \ #1} }

\newcommand{\complex}{{\mathbb C}} 
\newcommand{\nat}{{\mathbb N}} 
\newcommand{\real}{{\mathbb R}} 
\newcommand{\eucl}{{\mathbb E}}
\newcommand{\id}{{1\!\!1}} 

\newif\ifold             \oldtrue

\def\nn{\nonumber}

\def\e{{\,\rm e}\,}

\def\be{\begin{equation}}
\def\ee{\end{equation}}
\def\bea{\begin{eqnarray}}
\def\eea{\end{eqnarray}}
\def\bd{\begin{displaymath}}
\def\ed{\end{displaymath}}

\def\d{\partial}

\newcommand{\beq}{\begin{eqnarray}}
\newcommand{\eeq}{\end{eqnarray}}

\newcommand{\z}{\zeta}

\makeatletter
\newdimen\normalarrayskip              
\newdimen\minarrayskip                 
\normalarrayskip\baselineskip
\minarrayskip\jot
\newif\ifold             \oldtrue            
\def\arraymode{\ifold\relax\else\displaystyle\fi} 
\def\@arrayskip{\ifold\baselineskip\z@\lineskip\z@
     \else
     \baselineskip\minarrayskip\lineskip2\minarrayskip\fi}
\def\@arrayclassz{\ifcase \@lastchclass \@acolampacol \or
\@ampacol \or \or \or \@addamp \or
   \@acolampacol \or \@firstampfalse \@acol \fi
\edef\@preamble{\@preamble
  \ifcase \@chnum
     \hfil$\relax\arraymode\@sharp$\hfil
     \or $\relax\arraymode\@sharp$\hfil
     \or \hfil$\relax\arraymode\@sharp$\fi}}
\def\@array[#1]#2{\setbox\@arstrutbox=\hbox{\vrule
     height\arraystretch \ht\strutbox
     depth\arraystretch \dp\strutbox
     width\z@}\@mkpream{#2}\edef\@preamble{\halign \noexpand\@halignto
\bgroup \tabskip\z@ \@arstrut \@preamble \tabskip\z@ \cr}%
\let\@startpbox\@@startpbox \let\@endpbox\@@endpbox
  \if #1t\vtop \else \if#1b\vbox \else \vcenter \fi\fi
  \bgroup \let\par\relax
  \let\@sharp##\let\protect\relax
  \@arrayskip\@preamble}
\makeatother

\allowdisplaybreaks

\begin{document}
\begin{titlepage}
\begin{flushright}

\baselineskip=12pt

HWM--06--5\\
EMPG--06--02\\
hep--th/0602036\\
\hfill{ }\\
February 2006
\end{flushright}

\begin{center}

\vspace{2cm}

\baselineskip=24pt

{\Large\bf Noncommutative Field Theory \\ on Homogeneous Gravitational
  Waves}

\baselineskip=14pt

\vspace{1cm}

{\bf Sam Halliday} and {\bf Richard J. Szabo}
\\[4mm]
{\it Department of Mathematics}\\ and\\ {\it Maxwell Institute for
Mathematical Sciences\\ Heriot-Watt University\\ Colin Maclaurin Building,
  Riccarton, Edinburgh EH14 4AS, U.K.}
\\{\tt samuel@ma.hw.ac.uk} , {\tt R.J.Szabo@ma.hw.ac.uk}
\\[40mm]

\end{center}

\begin{abstract}

\baselineskip=12pt

We describe an algebraic approach to the time-dependent noncommutative
geometry of a six-dimensional Cahen-Wallach pp-wave string background
supported by a constant Neveu-Schwarz flux, and develop a general
formalism to construct and analyse quantum field theories defined
thereon. Various star-products are derived in closed explicit form and
the Hopf algebra of twisted isometries of the plane wave is
constructed. Scalar field theories are defined using explicit forms of
derivative operators, traces and noncommutative frame fields for the
geometry, and various physical features are described. Noncommutative
worldvolume field theories of D-branes in the pp-wave background are
also constructed.

\end{abstract}

\end{titlepage}
\setcounter{page}{2}

\newpage

\renewcommand{\thefootnote}{\arabic{footnote}} \setcounter{footnote}{0}

\newsection{Introduction and Summary\label{Intro}}

The general construction and analysis of noncommutative gauge theories
on curved spacetimes is one of the most important outstanding problems
in the applications of noncommutative geometry to string theory. These
non-local field theories arise naturally as certain decoupling limits
of open string dynamics on D-branes in curved superstring
backgrounds in the presence of a non-constant background Neveu-Schwarz
$B$-field. On a generic Poisson manifold $M$, they are formulated using
the Kontesevich star-product~\cite{Kont1} which is linked to a topological
string theory known as the Poisson sigma-model~\cite{CattFel1}. Under suitable
conditions, the quantization of D-branes in the Poisson sigma-model
which wrap coisotropic submanifolds of $M$, i.e. worldvolumes defined
by first-class constraints, may be consistently carried out and
related to the deformation quantization in the induced Poisson
bracket~\cite{CattFel2}. Branes defined by second-class constraints
may also be treated by quantizing Dirac brackets on the
worldvolumes~\cite{CFal1}.

However, in other concrete string theory settings, most studies of
noncommutative gauge theories on curved D-branes have been carried out
only within the context of the AdS/CFT correspondence by constructing
the branes as solutions in the dual supergravity description of the
gauge theory (see for example~\cite{Cai1,Cai2,HashSethi1,HashTh1,ASY1}). It is
important to understand how to describe the classical solutions and
quantization of these models directly at the field theoretic level in
order to better understand to what extent the noncommutative field
theories capture the non-local aspects of string theory and quantum
gravity, and also to be able to extend the descriptions to more
general situations which are not covered by the AdS/CFT
correspondence. In this paper we will investigate worldvolume
deformations in the simple example of the Hpp-wave background
$\NW_6$~\cite{Meessen1}, the six-dimensional Cahen-Wallach lorentzian
symmetric space $\CW_6$~\cite{CW1} supported by a constant null NS--NS
background three-form flux. The spacetime $\NW_6$ lifts to an exact
background of ten-dimensional superstring theory by taking the product
with an exact four-dimensional background, but we will not write this
explicitly. By projecting the transverse space of $\NW_6$ onto a plane
one obtains the four-dimensional Nappi-Witten spacetime
$\NW_4$~\cite{NW1}, and occasionally our discussion will pertain to
this latter exact string background. Our techniques are presented in a
manner which is applicable to a wider class of homogeneous pp-waves
supported by a constant Neveu-Schwarz flux.

Open string dynamics on this background is particularly
interesting because it has the potential to display a time-dependent
noncommutative geometry~\cite{DN1,HashSethi1}, and hence the
noncommutative field theories built on $\NW_6$ can serve as
interesting toy models for string cosmology which can be treated for
the most part as ordinary field theories. However, this point is
rather subtle for the present geometry~\cite{DN1,HashTh2}. A
particular gauge choice which leads to a time-dependent
noncommutativity parameter breaks conformal invariance of the
worldsheet sigma-model, i.e. it does not satisfy the Born-Infeld field
equations, while a conformally invariant background yields a
non-constant but time-independent noncommutativity. In this paper we
will partially clarify this issue. The more complicated noncommutative
geometry that we find contains both the transverse space dependent
noncommutativity between transverse and light-cone position
coordinates of the Hashimoto-Thomas model~\cite{HashTh2} and the
asymptotic time-dependent noncommutativity between transverse space
coordinates of the Dolan-Nappi model~\cite{DN1}.

The background $\NW_6$ arises as the Penrose-G\"uven
limit~\cite{Penrose1,Guven1} of an $\AdS_3\times\S^3$
background~\cite{BF-OFP1}. While this limit is a useful tool for
understanding various aspects of string dynamics, it is not in general
suitable for describing the quantum geometry of embedded
D-submanifolds~\cite{HSz1}. In the following we will resort to a more
direct quantization of the spacetime $\NW_6$ and its
D-submanifolds. We tackle the problem in a purely algebraic way by
developing the noncommutative geometry of the
universal enveloping algebra of the twisted Heisenberg algebra, whose
Lie group $\mathcal{N}$ coincides with the homogeneous spacetime $\CW_6$ in
question. While our algebraic approach has the advantage of yielding
very explicit constructions of noncommutative field theories in
these settings, it also has several limitations. It does not describe
the full quantization of the curved spacetime $\NW_6$, but rather only
the semi-classical limit of small NS--NS flux $\theta$ in which $\CW_6$
approaches flat six-dimensional Minkowski space. This is equivalent to
the limit of small light-cone time $x^+$ for the open string
dynamics. In this limit we can apply the Kontsevich formula to
quantize the pertinent Poisson geometry, and hence define
noncommutative worldvolume field theories of D-branes. Attempting to
quantize the full curved geometry (having $\theta\gg0$) would bring
us deep into the stringy regime~\cite{HoYeh1} wherein a field
theoretic analysis would not be possible. The worldvolume deformations
in this case are described by nonassociative algebras and variants of
quantum group algebras~\cite{LorCorn1,ARS1}, and there is no natural
notion of quantization for such geometries. We will nonetheless
emphasize how the effects of curvature manifest themselves in this
semi-classical limit.

The spacetime $\NW_6$ is wrapped by non-symmetric D5-branes which can
be obtained, as solutions of Type~II supergravity, from the
Penrose-G\"uven limit of spacetime-filling D5-branes in
$\AdS_3\times\S^3$~\cite{KNSanjay1}. This paper takes a very detailed
look at the first steps towards the construction and
analysis of noncommutative worldvolume field theories on these
branes. While we deal explicitly only with the case of scalar field
theory in detail, leaving the more subtle construction of
noncommutative gauge theory for future work, our results provide all
the necessary ingredients for analysing generic field theories in
these settings. We will also examine the problem of quantizing
regularly embedded D-submanifolds in $\NW_6$. The symmetric D-branes
wrapping twisted conjugacy classes of the Lie group $\mathcal{N}$ were
classified in~\cite{SF-OF1}. Their quantization was analysed
in~\cite{HSz1} and it was found that, in the semi-classical regime,
only the untwisted euclidean D3-branes support a noncommutative
worldvolume geometry. We study these D3-branes as a special case of
our more general constructions and find exact agreement with the
predictions of the boundary conformal field theory
analysis~\cite{DAK1}. We also find that the present technique captures
the noncommutative worldvolume geometry in a much more natural and
tractable way than the foliation of the group $\mathcal{N}$ by
quantized coadjoint orbits does~\cite{HSz1}. Our analysis is not
restricted to symmetric D-branes and can be applied to other
D-submanifolds of the spacetime $\NW_6$ as well.

The organisation of the remainder of this paper is as follows. In
Section~\ref{TTHA} we describe the twisted Heisenberg algebra, its
geometry, and the manner in which it may be quantized in the
semi-classical limit. In Section~\ref{StarProds} we construct
star-products which are equivalent to the Kontsevich product for the
pertinent Poisson geometry. These products are much simpler and more
tractable than the star-product on $\NW_6$ which was constructed
in~\cite{HSz1} through the noncommutative foliation of $\NW_6$ by
D3-branes corresponding to quantized coadjoint orbits. Throughout this
paper we will work with three natural star-products which we construct
explicitly in closed form. Two of them are canonically related to
coordinatizations of the classical pp-wave geometry, while the third
one is more natural from the algebraic point of view. We will derive
and compare our later results in all three of these star-product
deformations.

In Section~\ref{WeylSystems} we work out the corresponding generalized
Weyl systems~\cite{ALZ1} for these star-products, and use them in
Section~\ref{Coprod} to construct the Hopf algebras of twisted
isometries~\cite{CPT1,CKNT1,Wess1} of the noncommutative plane wave
geometry. In Section~\ref{Derivatives} we use the structure of this
Hopf algebra to build derivative operators. In contrast to more
conventional approaches~\cite{ConnesBook}, these operators are not
derivations of the star-products but are defined so that they are
consistent with the underlying noncommutative algebra of
functions. This ensures that the quantum group isometries, which carry
the nontrivial curvature content of the spacetime, act consistently on
the noncommutative geometry. In Section~\ref{Integrals} we define
integration of fields through a relatively broad class of consistent
traces on the noncommutative algebra of functions.

With these general constructions at hand, we proceed in
Section~\ref{FieldTheory} to analyse as a simple starting example the
case of free scalar field theory on the noncommutative spacetime
$\NW_6$. The analysis reveals the flat space limiting procedure in a
fairly drastic way. To get around this, we introduce noncommutative
frame fields which define derivations of the
star-products~\cite{BehrSyk1,HoMiao1}. Some potential physical
applications in the context of string dynamics in
$\NW_6$~\cite{DAK1,DAK2,BDAKZ1,CFS1,PK1} are also briefly
addressed. Finally, as another application we
consider in Section~\ref{D3Branes} the construction of noncommutative
worldvolume field theories of D-branes in $\NW_6$ using our general
formalism and compare with the quantization of symmetric D-branes
which was carried out in~\cite{HSz1}.

\newsection{Geometry of the Twisted Heisenberg Algebra\label{TTHA}}

In this section we will recall the algebraic definition~\cite{SF-OF1}
of the six-dimensional gravitational wave $\NW_6$ of Cahen-Wallach
type and describe the manner in which its geometry will be quantized
in the subsequent sections.

\subsection{Definitions \label{Defs}}

The spacetime NW$_6$ is defined as the group manifold of
the universal central extension of the subgroup
$\mathcal{S}:={\rm SO}(2)\ltimes\real^4$ of the four-dimensional
euclidean group ${\rm ISO}(4)={\rm
  SO}(4)\ltimes\real^4$. The corresponding simply connected group
$\mathcal N$ is homeomorphic to six-dimensional Minkowski space
$\eucl^{1,5}$. Its non-semisimple Lie algebra $\mathfrak n$ is
generated by elements $\J$, $\T$ and $\P^i_\pm$, $i=1,2$ obeying the
non-vanishing commutation relations
\bea
\left[\P^i_+\,,\,\P^j_-\right]&=&2\ii\delta^{ij}~\T \ , \nn\\
\left[\J\,,\,\P^i_\pm\right]&=&\pm\ii\P^i_\pm \ .
\label{NW4algdef}\eea
This is just the five-dimensional Heisenberg algebra extended by an
outer automorphism which rotates the noncommuting coordinates. The
twisted Heisenberg algebra
may be regarded as defining the harmonic oscillator algebra of a
particle moving in two dimensions, with the additional generator $\J$
playing the role of the number operator (or equivalently the
oscillator hamiltonian). It is this twisting that will
lead to a noncommutative geometry that deviates from the usual Moyal
noncommutativity generated by the Heisenberg algebra
(See~\cite{KS1,DougNek1,Sz1} for reviews in the present context). On the other
hand, $\mathfrak{n}$ is a solvable algebra whose properties are very
tractable. The subgroup $\mathcal{N}_0$ generated by $\P^1_\pm$, $\J$,
$\T$ is called the Nappi-Witten group and its four-dimensional group
manifold is the Nappi-Witten spacetime $\NW_4$~\cite{NW1}.

The most general invariant, non-degenerate symmetric bilinear form
$\langle-,-\rangle:\mathfrak{n}\times\mathfrak{n}\to\real$
is defined by the non-vanishing values~\cite{NW1}
\bea
\left\langle\P^i_+\,,\,\P^j_-\right\rangle&=&2\,\delta^{ij} \ , \nn\\
\left\langle\J\,,\,\T\right\rangle&=&1 \ , \nn\\
\left\langle\J\,,\,\J\right\rangle&=&b \ .
\label{NW4innerprod}\eea
The arbitrary parameter $b\in\real$ can be set to zero by a Lie
algebra automorphism of $\mathfrak{n}$. This inner product has
Minkowski signature, so that the
group manifold of $\mathcal N$ possesses a
homogeneous, bi-invariant lorentzian metric defined by the pairing of
the Cartan-Maurer left-invariant $\mathfrak n$-valued one-forms
$g^{-1}~\dd g$ for $g\in\mathcal N$ as
\beq
\dd s^2=\left\langle g^{-1}~\dd g\,,\,g^{-1}~\dd g\right\rangle \ .
\label{NW4CM}\eeq
A generic group element $g\in\mathcal N$ may be parametrized as
\beq
g(u,v,\ma,\overline{\ma}\,)=\e^{a_i\,\P^i_++\overline{a}_i
\,\P^i_-}~\e^{\theta\,u\,\J}~\e^{\theta^{-1}\,v\,\T}
\label{NW4coords}\eeq
with $u,v,\theta\in\real$ and $\ma=(a_1,a_2)\in\complex^2$. In these
global coordinates, the metric (\ref{NW4CM}) reads
\beq
\dd s^2=2~\dd u~\dd v+\dd\ma\cdot\dd\overline{\ma}+2\ii\theta\,\left(\ma\cdot
\dd\overline{\ma}-\overline{\ma}\cdot\dd\ma\right)~\dd u \ .
\label{NW4metricNW}\eeq
The metric (\ref{NW4metricNW}) assumes the standard form of the plane
wave metric of a conformally flat, indecomposable Cahen-Wallach
lorentzian symmetric spacetime CW$_6$ in six dimensions~\cite{CW1} upon
introduction of Brinkman coordinates~\cite{Brinkman1} $(x^+,x^-,\mz)$
defined by rotating the transverse space at a Larmor frequency as $u=x^+$,
$v=x^-$ and $\ma=\e^{\ii\theta\,x^+/2}\,\mz$. In these coordinates the
metric assumes the stationary form
\beq
\dd s^2=2~\dd x^+~\dd x^-+\dd\mz\cdot\dd\overline{\mz}
-\mbox{$\frac14$}\,\theta^2\,|\mz|^2~
\left(\dd x^+\right)^2 \ ,
\label{NW4metricBrink}\eeq
revealing the pp-wave nature of the geometry. Note that on the null
hyperplanes of constant $u=x^+$, the geometry becomes that of
flat four-dimensional euclidean space $\eucl^4$. This is the geometry
appropriate to the Heisenberg subgroup of $\mathcal{N}$, and is what
is expected in the Moyal limit when the effects of the extra generator
$\J$ are turned off.

The spacetime NW$_6$ is further supported by a Neveu-Schwarz two-form
field $B$ of constant field strength
\bea
H&=&-\mbox{$\frac13$}\,\bigl\langle g^{-1}~\dd g\,,\,\dd
\left(g^{-1}~\dd g\right)\bigl\rangle
{}~=~2\ii\theta~\dd x^+\wedge\dd\mz^\top\wedge\dd\overline{\mz}
{}~=~\dd B \ , \nn\\
B&=&-\mbox{$\frac12$}\,\bigl\langle g^{-1}~\dd g\,,\,
\frac{\id+{\rm Ad}_g}{\id-{\rm Ad}_g}\,g^{-1}~\dd g\bigl\rangle~=~
2\ii\theta\,x^+~\dd\mz^\top\wedge\dd\overline{\mz} \ ,
\label{NS2formBrink}\eea
defined to be non-vanishing only on vectors tangent to the conjugacy
class containing $g\in\mathcal{N}$~\cite{AlekSch1}. It is the presence of
this $B$-field that induces time-dependent noncommutativity of the
string background in the presence of D-branes. Because its flux is
constant, the noncommutative dynamics in certain kinematical regimes
on this space can still be formulated exactly, just like on other
symmetric curved noncommutative spaces (See~\cite{Schomrev} for a
review of these constructions in the case of compact group
manifolds).

\subsection{Quantization\label{NWQuant}}

We will now begin working our way towards describing how
the worldvolumes of D-branes in the spacetime $\NW_6$ are
deformed by the non-trivial $B$-field background. The Seiberg-Witten
bi-vector~\cite{SW1} induced by the Neveu-Schwarz background
(\ref{NS2formBrink}) and the pp-wave metric $G$ given by
(\ref{NW4metricBrink}) is
\beq
\Theta=-(G+B)^{-1}\,B\,(G-B) \ .
\label{SWTheta}\eeq
Let us introduce the one-form
\beq
\Lambda:=-\ii\left(\theta^{-1}\,x_0^-+\theta\,x^+\right)\,\left(
\mz\cdot\dd\overline{\mz}-\overline{\mz}\cdot\dd\mz
\right)
\label{Lambdadef}\eeq
on the null hypersurfaces of constant $x^-=x_0^-$, and compute the
corresponding two-form gauge transformation of the $B$-field in
(\ref{NS2formBrink}) to get
\beq
B~\longmapsto~B+\dd\Lambda=-\ii\theta~\dd x^+\wedge\left(
\mz\cdot\dd\overline{\mz}-\overline{\mz}\cdot\dd\mz
\right)+2\ii\theta\,x_0^-~\dd\overline{\mz}{}^{\,\top}
\wedge\dd\mz \ .
\label{B6gaugeequiv}\eeq
The Seiberg-Witten bi-vector in this gauge is given by~\cite{HSz1}
\beq
\Theta=-\mbox{$\frac{2\ii\theta}{\theta^2+\left(x_0^-\right)^2}$}\,
\left[\theta^2~\partial_-\wedge\left(\mz\cdot\mdell-
\overline{\mz}\cdot\overline{\mdell}\,\right)+4x_0^-~
\mdell^\top\wedge\overline{\mdell}\,\right] \ ,
\label{ThetaLambda}\eeq
where $\partial_\pm:=\frac\partial{\partial x^\pm}$ and
$\mdell=(\partial^1,\partial^2):=(\frac\partial{\partial
  z_1},\frac\partial{\partial z_2})$. Since (\ref{ThetaLambda}) is
degenerate on the whole $\NW_6$ spacetime, it does not define a
symplectic structure. However, one easily checks that it does define a
Poisson structure, i.e. $\Theta$ is a Poisson
bi-vector~\cite{HSz1}. In this gauge one can show that a consistent
solution to the Born-Infeld equations of motion on a non-symmetric
spacetime-filling D5-brane wrapping $\NW_6$ has vanishing ${\rm U}(1)$
gauge field flux $F=0$~\cite{HashTh2}.

In particular, at the special value $x_0^-=\theta$ and with the
rescaling $\mz\to\sqrt{2/\theta\,\tau}~\mz$, the corresponding open
string metric~\cite{SW1} $G_{\rm open}=G-B\,G^{-1}\,B$ becomes that of
$\CW_6$ in global coordinates (\ref{NW4metricNW})~\cite{HSz1}, while
the non-vanishing Poisson brackets corresponding to
(\ref{ThetaLambda}) read
\beq
\left\{z_i\,,\,\overline{z}_j\right\}&=&2\ii\theta\,\tau~\delta_{ij} \ ,
\nonumber\\ \left\{x^-\,,\,z_i\right\}&=&-\ii\theta\,z_i \ ,
\nonumber\\ \left\{x^-\,,\,\overline{z}_i\right\}&=&\ii\theta\,
\overline{z}_i
\label{Poissonspecial}\eeq
for $i,j=1,2$. The Poisson algebra thereby coincides with the
Lie algebra $\mathfrak{n}$ in this case and the metric
on the branes with the standard curved geometry of the pp-wave. In the
semi-classical flat space limit $\theta\to0$, the
quantization of the brackets (\ref{Poissonspecial}) thereby yields a
noncommutative worldvolume geometry on D5-branes wrapping $\NW_6$
which can be associated with a quantization of $\mathfrak{n}$ (or
more precisely of its dual $\mathfrak{n}^{\vee\,}$). In this limit, the
corresponding quantization of $\NW_6$ is thus given by the associative
Kontsevich star-product~\cite{Kont1}. Henceforth, with a slight abuse
of notation, we will denote the central coordinate $\tau$ as the plane
wave time coordinate $x^+$. Our semi-classical quantization will then
be valid in the small time limit $x^+\to0$.

Our starting point in describing the noncommutative geometry of
$\NW_6$ will therefore be at the algebraic level. We will consider the
deformation quantization of the dual $\mathfrak{n}^{\vee\,}$ to the Lie
algebra $\mathfrak{n}$. Naively,
one may think that the easiest way to carry this out is to compute star
products on the pp-wave by taking the Penrose limits of the
standard ones on $\S^3$ and $\AdS_3$ (or equivalently by contracting
the standard quantizations of the Lie algebras ${\rm su}(2)$ and ${\rm
  sl}(2,\real)$). However, some quick calculations show that the
induced star-products obtained in this way are divergent in the
infinite volume limit, and the reason why is
simple. While the standard In\"on\"u-Wigner contractions hold at the
level of the Lie algebras~\cite{SF-OF1}, they need not necessarily map
the corresponding universal enveloping algebras, on
which the quantizations are performed. This is connected to the
phenomenon that twisted conjugacy classes of branes are not
necessarily related by the Penrose-G\"uven limit~\cite{HSz1}. We must
therefore resort to a more direct approach to quantizing the spacetime
$\NW_6$.

For notational ease, we will write the algebra $\mathfrak{n}$ in the
generic form
\beq
[\X_a,\X_b]=\ii\theta\,C_{ab}^{~~c}\,\X_c \ ,
\label{n6genform}\eeq
where $(\X_a):=\theta\,(\J,\T,\P^i_\pm)$ are the
generators of $\mathfrak{n}$ and the structure constants
$C_{ab}^{~~c}$ can be gleamed off from (\ref{NW4algdef}). The algebra
(\ref{n6genform}) can be regarded
as a formal deformation quantization of the Kirillov-Kostant Poisson bracket on
$\mathfrak{n}^{\vee\,}$ in the standard coadjoint orbit method. Let us
identify $\mathfrak{n}^{\vee\,}$ as the vector space $\real^6$ with basis
$\X_a^{\vee\,}:=\langle\X^{~}_a,-\rangle:\mfn\to\real$ dual to the
$\X_a^{~}$. In the algebra of polynomial functions
$\complex(\mathfrak{n}^{\vee\,})=\complex(\real^6)$, we may then identify
the generators $\X_a$ themselves with the coordinate functions
\bea
\X^{~}_\J(\mx)&=&x^{~}_\T~=~x^- \ , \nn\\ \X^{~}_\T(\mx)
&=&x^{~}_\J~=~x^+ \ , \nn\\
\X^{~}_{\P_+^i}(\mx)&=&2x^{~}_{\P_-^i}~=~2\overline{z}_i \ ,
\nn\\ \X^{~}_{\P_-^i}(\mx)&=&2x^{~}_{\P_+^i}~=~2z_i
\label{Xacoordfns}\eea
for any $\mx\in\mathfrak{n}^{\vee\,}$ with component $x_a$ in the
$\X_a^{\vee\,}$
direction. These functions generate the whole coordinate algebra and
their Poisson bracket $\Theta$ is defined by
\beq
\Theta(\X_a,\X_b)(\mx)=\mx\bigl([\X_a,\X_b]\bigr) ~~~~ \forall\mx\in
\mathfrak{n}^{\vee\,} \ .
\label{KKXadef}\eeq
Therefore, when viewed as functions on $\real^6$ the Lie algebra
generators have a Poisson bracket given by the Lie bracket, and their
quantization is provided by (\ref{n6genform}) with deformation
parameter~$\theta$. In the next section
we will explore various aspects of this quantization and derive
several (equivalent) star products on $\mfn^{\vee\,}$.

\newsection{Gutt Products\label{StarProds}}

The formal completion of the space of polynomials
$\complex(\mfn^{\vee\,})$ is the algebra ${\rm
  C}^\infty(\mathfrak{n}^{\vee\,})$ of smooth functions on
$\mathfrak{n}^{\vee\,}$. There is a natural way to construct a
star-product on the cotangent bundle
$T^*\mathcal{N}\cong\mathcal{N}\times\mathfrak{n}^{\vee\,}$, which
naturally induces an associative product on ${\rm
  C}^\infty(\mfn^{\vee\,})$. This induced product is called the Gutt
product~\cite{Gutt1}. The Poisson bracket defined by (\ref{KKXadef})
naturally extends to a Poisson structure
$\Theta:\CC^\infty(\mfn^{\vee\,})\times\CC^\infty(\mfn^{\vee\,})
\to\CC^\infty(\mfn^{\vee\,})$ defined by the Kirillov-Kostant
bi-vector
\beq
\Theta=\mbox{$\frac12$}\,C_{ab}^{~~c}\,x_c\,\partial^a\wedge\partial^b
\label{KKbivector}\eeq
where $\partial^a:=\frac{\partial}{\partial x_a}$. This coincides with
the Seiberg-Witten bi-vector in the limits described in
Section~\ref{NWQuant}. The Gutt product constructs a
  quantization of this Poisson structure. It is equivalent to the
  Kontsevich star-product in this case~\cite{Dito1}, and by
  construction it keeps that part of the Kontsevich formula which is
  associative~\cite{Shoikhet1}. In general, within the present
  context, the Gutt and Kontsevich deformation quantizations are only
  identical for nilpotent Lie algebras~\cite{Kathotia1}.

The algebra $\complex(\mfn^{\vee\,})$ of polynomial functions on the dual
to the Lie algebra is naturally isomorphic to the symmetric tensor algebra
$S(\mfn)$ of $\mfn$. By the Poincar\'e-Birkhoff-Witt theorem,
there is a natural isomorphism $\Omega:S(\mfn)\to
U(\mfn)$ with the universal enveloping algebra $U(\mfn)$ of
$\mfn$. Using the above identifications, this extends to a canonical
isomorphism
\beq
\Omega\,:\,\CC^\infty\left(\real^6\right)~\longrightarrow~\overline{
U(\mfn)^\complex}
\label{Sigmaiso}\eeq
defined by specifying an ordering for the elements of the
basis of monomials for $S(\mfn)$, where
$\overline{U(\mfn)^\complex}$ denotes a formal completion of the
complexified universal enveloping algebra
$U(\mathfrak{n})^\complex:=U(\mathfrak{n})\otimes\complex$.
Denoting this ordering by $\NO-\NO$, we may write this
isomorphism symbolically as
\beq
\Omega(x_{a_1}\cdots x_{a_n})=\NO\,\X_{a_1}\cdots\X_{a_n}\,\NO \ .
\label{Sigmasymbol}\eeq
The original Gutt construction~\cite{Gutt1} takes the isomorphism
$\Omega$ on $S(\mfn)$ to be symmetrization of monomials. In this
case $\Omega(f)$ is usually called the Weyl symbol of
$f\in\CC^\infty(\real^6)$ and the symmetric ordering $\NO-\NO$
of symbols $\Omega(f)$ is called Weyl ordering. In the following we
shall work with three natural orderings appropriate to the
algebra $\mfn$.

The isomorphism (\ref{Sigmaiso}) can be used to transport the
algebraic structure on the universal enveloping algebra $U(\mfn)$ of
$\mfn$ to the algebra of smooth functions on $\mfn^{\vee\,}\cong\real^6$
and give the star-product defined by
\beq
f\star g:=\Omega^{-1}\bigl(\,\NO\,\Omega(f)\cdot\Omega(g)\,\NO
\,\bigr) \ , ~~ f,g\in\CC^\infty\left(\real^6\right) \ .
\label{fstargSigma}\eeq
The product on the right-hand side of the formula (\ref{fstargSigma})
is taken in $U(\mfn)$, and it follows that $\star$ defines an
associative, noncommutative product. Moreover, it represents a
deformation quantization of the Kirillov-Kostant Poisson structure on
$\mfn^{\vee\,}$, in the sense that
\beq
[x,y]_\star:=x\star y-y\star x=\ii
\theta\,\Theta(x,y) \ , ~~ x,y\in\complex_{(1)}\left(\mfn^{\vee\,}\right) \ ,
\label{xyPoisson}\eeq
where $\complex_{(1)}(\mfn^{\vee\,})$ is the subspace of homogeneous
polynomials of degree~$1$ on $\mfn^{\vee\,}$. In particular, the Lie
algebra relations (\ref{n6genform}) are reproduced by star-commutators
of the coordinate functions as
\beq
[x_a,x_b]_\star=\ii\theta\,C_{ab}^{~~c}\,x_c \ ,
\label{xaxbstarcomm}\eeq
in accordance with the Poisson brackets (\ref{Poissonspecial}) and the
definition (\ref{KKXadef}).

Let us now describe how to write the star-product (\ref{fstargSigma})
explicitly in terms of a bi-differential operator
$\hat{\mathcal{D}}:\CC^\infty(\mfn^{\vee\,})\times\CC^\infty(\mfn^{\vee\,})\to
\CC^\infty(\mfn^{\vee\,})$~\cite{Kathotia1}. Using the
Kirillov-Kostant Poisson structure as
before, we identify the generators of $\mfn$ as coordinates on
$\mfn^{\vee\,}$. This establishes, for small $s\in\real$, a one-to-one
correspondence between group elements $\e^{s\,\X}$, $\X\in\mfn$ and
functions $\e^{s\,x}$ on $\mfn^{\vee\,}$. Pulling back the group
multiplication of elements $\e^{s\,\X}\in\mathcal{N}$ via this
correspondence induces a bi-differential operator $\hat{\mathcal{D}}$
acting on the functions $\e^{s\,x}$. Since these functions separate
the points on $\mfn^{\vee\,}$, this extends to an operator on the whole
of $\CC^\infty(\mfn^{\vee\,})$.

To apply this construction explicitly, we use the following
trick~\cite{MSSW1,BehrSyk1} which will also prove useful for later
considerations. By restricting to an appropriate Schwartz subspace of
functions $f\in\CC^\infty(\real^6)$, we may use a Fourier
representation
\beq
f(\mx)=\int\limits_{\real^6}\frac{\dd\mk}{(2\pi)^6}~\tilde f(\mk)~
\e^{\ii\mk\cdot\mx} \ .
\label{Fouriertransfdef}\eeq
This establishes a correspondence between (Schwartz) functions on
$\mfn^{\vee\,}$ and elements of the complexified group
$\mathcal{N}^\complex:=\mathcal{N}\otimes\complex$. The products
of symbols $\Omega(f)$ may be computed using (\ref{Sigmasymbol}), and
the star-product (\ref{fstargSigma}) can be represented in terms of a
product of group elements in $\mathcal{N}^\complex$ as
\beq
f\star g=\int\limits_{\real^6}\frac{\dd\mk}{(2\pi)^6}~
\int\limits_{\real^6}\frac{\dd\mq}{(2\pi)^6}~\tilde f(\mk)\,
\tilde g(\mq)~\Omega^{-1}\left(\,\NO~~\NO\,\e^{\ii k^a\,\X_a}\,\NO\cdot
\NO\,\e^{\ii q^a\,\X_a}\,\NO~~\NO\,\right) \ .
\label{fstargFourier}\eeq
Using the Baker-Campbell-Hausdorff formula, to be discussed below, we
may write
\beq
\NO~~\NO\,\e^{\ii k^a\,\X_a}\,\NO\cdot\NO\,\e^{\ii q^a\,\X_a}\,\NO~~\NO=
\NO\,\e^{\ii D^a(\mk,\mq)\,\X_a}\,\NO
\label{NOproductsBCH}\eeq
for some function $\mD=(D^a):\real^6\times\real^6\to\real^6$. This
enables us to rewrite the star-product (\ref{fstargFourier}) in terms
of a bi-differential operator $f\star g:=\hat{\mathcal{D}}(f,g)$ given
explicitly by
\beq
f\star
g=f~\e^{\ii\mx\cdot[\mD(\,-\ii\overleftarrow{\mdell}
\,,\,-\ii\overrightarrow{\mdell}\,)+\ii\overleftarrow{\mdell}+\ii
\overrightarrow{\mdell}\,]}~g
\label{fstargbidiff}\eeq
with $\mdell:=(\partial^a)$. In particular, the star-products of
the coordinate functions themselves may be computed from the formula
\beq
x_a\star x_b=\left.-\frac{\partial}{\partial k^a}\frac\partial
{\partial q^b}\e^{\ii\mD(\mk,\mq)\cdot\mx}\right|_{\mk=\mq=\mbf0} \ .
\label{xastarxb}\eeq

Finally, let us describe how to explicitly compute the functions
$D^a(\mk,\mq)$ in (\ref{NOproductsBCH}). For this, we consider the
Dynkin form of the Baker-Campbell-Hausdorff formula which is given for
$\X,\Y\in\mfn$ by
\begin{equation}
  \label{eq:BCH:define}
\e^\X~\e^\Y=\e^{\mathrm{H}(\X:\Y)} \ ,
\end{equation}
where $\mathrm{H}(\X:\Y)=\sum_{n\geq1}\mathrm{H}_n(\X:\Y)\in\mfn$ is
generically an infinite series whose terms may be calculated through the
recurrence relation
\begin{eqnarray}
  \label{eq:BCH}
&&(n+1)~\mathrm{H}_{n+1}(\X:\Y)~=~\mbox{$\frac 12$}\,\bigl[\X-\Y\,,\,
\mathrm{H}_n(\X:\Y)\bigr]
\nonumber  \\ &&~~~~~~~~~~~~~~~~~~~~
+\,\sum_{p=1}^{\lfloor n/2\rfloor}\frac{B_{2p}}{(2p)!}~
\sum_{\substack{k_1,\ldots,k_{2p}> 0 \\ k_1+\ldots+k_{2p}=n }}
\bigl[\mathrm{H}_{k_1}(\X:\Y)\,,\,\bigl[\,\ldots\,,\,\bigl[
\mathrm{H}_{k_{2p}}(\X:\Y)\,,\,\X+\Y\bigr]\ldots\bigr]\,\bigr]\nonumber\\
\end{eqnarray}
with $\mathrm{H}_1(\X:\Y):=\X+\Y$. The coefficients $B_{2p}$ are the
Bernoulli numbers which are defined by the generating function
\begin{equation}
  \label{eq:BCH:K}
  \frac{s}{1-\e^{-s}}-\frac s2-1=\sum_{p=1}^\infty\frac{B_{2p}}{(2p)!}
  ~s^{2p} \ .
\end{equation}
The first few terms of the formula (\ref{eq:BCH:define}) may be
written explicitly as
\begin{eqnarray}
  \label{eq:BCH:1}
  \mathrm{H}_1(\X:\Y)&=& \X+\Y \ , \nonumber\\
  \mathrm{H}_2(\X:\Y)&=&\mbox{$\frac 12$}\,\cb \X\Y \ , \nonumber \\
  \mathrm{H}_3(\X:\Y)&=&\mbox{$\frac 1{12}$}\,\bigl[\X\,,\,\cb \X\Y\,\bigr]
  -\mbox{$\frac 1{12}$}\,\bigl[\Y\,,\,\cb \X\Y\,\bigr] \ , \nonumber\\
  \mathrm{H}_4(\X:\Y)&=& -\mbox{$\frac 1{24}$}\,\bigl[\Y\,,\,\bigl[\X
  \,,\,\cb \X\Y\,\bigr]\,\bigr] \ .
\end{eqnarray}
Terms in the series grow increasingly complicated due to
the sum over partitions in \eqref{eq:BCH}, and in general there is no
closed symbolic form, as in the case of the Moyal product based on the
ordinary Heisenberg algebra, for the functions $D^a(\mk,\mq)$ appearing in
(\ref{NOproductsBCH}). However, at least for certain ordering
prescriptions, the solvability of the Lie algebra $\mfn$ enables one
to find explicit expressions for the star-product (\ref{fstargbidiff})
in this fashion. We will now proceed to construct three such
products.

\subsection{Time Ordering\label{TOP}}

The simplest Gutt product is obtained by choosing a ``time ordering''
prescription in (\ref{Sigmasymbol}) whereby all factors of the time
translation generator $\J$ occur to the far right in any monomial in
$U(\mfn)$. It coincides precisely with the global coordinatization
(\ref{NW4coords}) of the Cahen-Wallach spacetime, and written on
elements of the complexified group
$\mathcal{N}^\complex$ it is defined by
\begin{equation}
  \label{eq:time:defn}
\Omega_*\left(\e^{\ii\mk\cdot\mx}\right)=
\NOa\,\e^{\ii k^a\,\X_a}\,\NOa:=\e^{\ii(p_i^+\,\P^i_+
+p_i^-\,\P^i_-)}~\e^{\ii j\,\J}~\e^{\ii t\,\T} \ ,
\end{equation}
where we have denoted
$\mk:=(j,t,\mbp^\pm)$ with $j,t\in\real$ and
$\mbp^\pm=\overline{\mbp^\mp}=(p_1^\pm,p_2^\pm)\in\complex^2$. To
calculate the corresponding star-product $*$, we have to compute the
group products
\bea
\NOa~~\NOa\,\e^{\ii k^a\,\X_a}\,\NOa\cdot\NOa\,\e^{\ii k^{\prime\,a}\,\X_a}
\,\NOa~~\NOa&=&\NOa\,\e^{\ii(p_i^+\,\P^i_++p_i^-\,\P^i_-)}~\e^{\ii j\,\J}~
\e^{\ii t\,\T}\nonumber\\ && \quad\times~
\e^{\ii(p_i^{\prime\,+}\,\P^i_++p_i^{\prime\,-}\,\P^i_-)}~\e^{\ii j'\,\J}~
\e^{\ii t'\,\T}\,\NOa \ .
\label{TOgpprods}\eea

The simplest way to compute these products is to realize the
six-dimensional Lie algebra $\mfn$ as a central
extension of the subalgebra $\mfs={\rm so}(2)\ltimes\real^4$ of the
four-dimensional euclidean algebra
${\rm iso}(4)={\rm
  so}(4)\ltimes\real^4$~\cite{SF-OF1,F-OFS1}. Regarding $\real^4$ as
$\complex^2$ (with respect to a chosen complex structure), for generic
$\theta\neq0$ the generators of $\mfn$ act on $\mw\in\complex^2$
according to the affine transformations $\e^{\ii
  j\,\J}\cdot\mw=\e^{-\theta\,j}\,\mw$ and
$\e^{\ii(p_i^+\,\P^i_++p_i^-\,\P^i_-)}\cdot\mw=\mw+\ii\theta\,\mbp^-$,
corresponding to a combined rotation in the $(12)$, $(34)$ planes and
translations in $\real^4\cong\complex^2$. The central element
generates an abstract one-parameter subgroup acting as $\e^{\ii
  t\,\T}\cdot\mw=\e^{-\theta\,t}\,\mw$ in this representation. From
this action we can read off the group multiplication laws
\bea
\e^{\ii j\,\J}~\e^{\ii j'\,\J}&=&\e^{\ii(j+j'\,)\,\J} \ ,
\label{JJgpmultlaw}\\
{~~~~}^{~~}_{~~}\nn\\\e^{\ii
  j\,\J}~\e^{\ii(p_i^+\,\P^i_++p_i^-\,\P^i_-)}
&=&\e^{\ii(\,\e^{-\theta\,j}\,p_i^+\,\P^i_+
+\e^{\theta\,j}\,p_i^-\,\P^i_-)}~\e^{\ii j\,\J} \ ,
\label{JQgpmultlaw}\\
{~~~~}^{~~}_{~~}\nn\\\e^{\ii(p_i^+\,\P^i_++p_i^-\,\P^i_-)}~
\e^{\ii(p_i^{\prime\,+}\,\P^i_++p_i^{\prime\,-}\,\P^i_-)}~
&=&\e^{\ii[(p_i^++p_i^{\prime\,+})\,\P^i_+
+(p_i^-+p_i^{\prime\,-})\,\P^i_-]}~\e^{2\theta~{\rm Im}(
\mbp^+\cdot\mbp^{\prime\,-})\,\T}
\label{QQgpmultlaw}\eea
where the formula (\ref{JQgpmultlaw}) displays the semi-direct product
nature of the euclidean group, while (\ref{QQgpmultlaw}) displays the
group cocycle of the projective representation of the subgroup
$\mathcal S$ of ${\rm ISO}(4)$, arising from the central extension,
which makes the translation algebra noncommutative and is computed
from the Baker-Campbell-Hausdorff formula.

Using (\ref{JJgpmultlaw})--(\ref{QQgpmultlaw}) we may now compute the
products (\ref{TOgpprods}) and one finds
\bea
\NOa~~\NOa\,\e^{\ii k^a\,\X_a}\,\NOa\cdot\NOa\,\e^{\ii k^{\prime\,a}\,\X_a}
\,\NOa~~\NOa&=&\e^{\ii[(p_i^++\e^{-\theta\,j}\,p_i^{\prime\,+})\,
\P^i_++(p_i^-+\e^{\theta\,j}\,p_i^{\prime\,-})\,\P^i_-]}~
\e^{\ii(j+j'\,)\,\J}\nonumber\\ &&\times~
\e^{\ii[t+t'-\theta\,(\e^{\theta\,j}\,
\mbp^+\cdot\mbp^{\prime\,-}-\e^{-\theta\,j}\,
\mbp^-\cdot\mbp^{\prime\,+})]\,\T} \ .
\label{TOgpprodexpl}\eea
{}From (\ref{xastarxb}) we may compute the star-products between the
coordinate functions on $\mfn^{\vee\,}$ and easily verify the commutation
relations of the algebra $\mfn$,
\bea
x_a*x_a&=&(x_a)^2 \ , \nonumber\\x_a*x^+&=&x^+*x_a~=~x_a\,x^+ \ ,
\nonumber\\z_1*z_2&=&z_2*z_1~=~z_1\,z_2 \ , \nonumber\\
\overline{z}_1*\overline{z}_2&=&\overline{z}_2*\overline{z}_1
{}~=~\overline{z}_1\,\overline{z}_2 \ , \nonumber\\
x^-*z_i&=&x^-\,z_i-\ii\theta\,z_i \ , \nonumber\\
z_i*x^-&=&x^-\,z_i \ , \nonumber\\
x^-*\overline{z}_i&=&x^-\,\overline{z}_i+\ii\theta\,\overline{z}_i \ ,
\nonumber\\ \overline{z}_i*x^-&=&x^-\,\overline{z}_i \ , \nonumber\\
z_i*\overline{z}_i&=&z_i\,\overline{z}_i-\ii\theta\,x^+ \ , \nn\\
\overline{z}_i*z_i&=&z_i\,\overline{z}_i+\ii\theta\,x^+ \ ,
\label{TOcoordstarprods}\eea
with $a=1,\dots,6$ and $i=1,2$. From
(\ref{NOproductsBCH},\ref{fstargbidiff}) we find the star-product $*$
of generic functions $f,g\in\CC^{\infty}(\mfn^{\vee\,})$ given by
\bea
f*g&=&\mu\circ\exp\left[\ii\theta\,x^+\,\left(\e^{-\ii\theta\,\partial_-}\,
\mdell^\top\otimes\overline{\mdell}-\e^{\ii\theta\,\partial_-}\,
{\overline{\mdell}}{}^{\,\top}\otimes\mdell\right)\right.\nonumber
\\ &&\qquad\qquad+\left.\overline{z}_i\,
\left(\e^{\ii\theta\,\partial_-}-1\right)\otimes
\partial^i+z_i\,\left(\e^{-\ii\theta\,\partial_-}-1\right)
\otimes\overline{\partial}{}^{\,i}\right]f\otimes g \ ,
\label{TOstargen}\eea
where $\mu(f\otimes g)=f\,g$ is the pointwise product. To second order
in the deformation parameter $\theta$ we obtain
\begin{eqnarray}
  \label{eq:time:positionspace}
  \nonumber
  f\ast g&=&f\,g
  -\ii\theta\,\left[
    x^+\,\left(\,\overline{\mdell}f\cdot\mdell g
    -\mdell f\cdot\overline{\mdell}g\right)
    -\overline{\mz}\cdot\d_-f\,\mdell g
    +\mz\cdot\d_-f\,\overline{\mdell}g
  \right]\\\nonumber
  &&-\,\theta^2\,\mbox{$\sum\limits_{i=1,2}$}\,\left[\,
  \mbox{$\frac12$}\,\left(x^+\right)^2\,\left((\d^i)^2f\,
  (\,{\overline{\d}}{}^{\,i})^2g
  -2{\overline{\d}}{}^{\,i}\d^if\,{\overline{\d}}{}^{\,i}\d^ig
  +({\overline{\d}}{}^{\,i})^2f\,(\d^i)^2g\right)\right.
  \\\nonumber&&\qquad\qquad\quad-\,x^+\,\left(\d^i\d_- f\,
  {\overline{\d}}{}^{\,i}g
  -{\overline{\d}}{}^{\,i}\d_- f\,\d^ig\right)
  -x^+\,\overline{z}_i\,\left(\,{\overline{\d}}{}^{\,i}\d_- f\,(\d^i)^2g
  -\d^i\d_- f\,{\overline{\d}}{}^{\,i}\d^ig\right)\\
  \nonumber &&\qquad\qquad\quad
  +\,x^+\,z_i\,\left(\,{\overline{\d}}{}^{\,i}\d_-
    f\,{\overline{\d}}{}^{\,i}\d^ig
  -\d^i\d_- f\,(\,{\overline{\d}}{}^{\,i})^2g\right)
  -\overline{z}_i\,z_i\,\d_-^2f\,{\overline{\d}}{}^{\,i}\d^ig
  \\ \nonumber &&\qquad\qquad\quad
  +\Bigl.\mbox{$\frac12$}\,\left(\overline{z}_i^{\,2}\,\d_-^2f\,(\d^i)^2g
  +\overline{z}_i\,\d_-^2f\,\d^ig
  +z_i\,\d_-^2f\,{\overline{\d}}{}^{\,i}g
  +z_i^2\,\d_-^2f\,(\,{\overline{\d}}{}^{\,i})^2g\right)
  \Bigr]\\ && +\,O\left(\theta^3\right) \ .
\end{eqnarray}

\subsection{Symmetric Time Ordering\label{TSOP}}

Our next Gutt product is obtained by taking a ``symmetric time
ordering'' whereby any monomial in $U(\mfn)$ is the symmetric sum
over the two time orderings obtained by placing $\J$ to the far right
and to the far left. This ordering is induced by the group contraction of
${\rm U}(1)\times{\rm SU}(2)$ onto the Nappi-Witten group
$\mathcal{N}_0$~\cite{DAK2}, and it thereby induces the
coordinatization of $\NW_4$ that is obtained from the
Penrose-G\"uven limit of the spacetime $\S^{1,0}\times\S^3$, i.e. it
coincides with the Brinkman coordinatization of the Cahen-Wallach
spacetime. On elements of $\mathcal{N}^\complex$ it is defined by
\beq
\Omega_\bullet\left(\e^{\ii\mk\cdot\mx}\right)=
\NOb\,\e^{\ii k^a\,\X_a}\,\NOb:=\e^{\frac\ii2\,j\,\J}~
\e^{\ii(p_i^+\,\P^i_+
+p_i^-\,\P^i_-)}~\e^{\frac\ii2\,j\,\J}~\e^{\ii t\,\T} \ .
\label{TOsymgpprods}\eeq
{}From (\ref{JJgpmultlaw})--(\ref{QQgpmultlaw}) we can again easily
compute the required group products to get
\bea
\NOb~~\NOb\,\e^{\ii k^a\,\X_a}\,\NOb\cdot\NOb\,\e^{\ii k^{\prime\,a}\,\X_a}
\,\NOb~~\NOb&=&\e^{\frac\ii2\,(j+j'\,)\,\J}\nonumber\\ &&\times~
\e^{\ii[(\e^{\frac{\theta}2\,j'}\,p_i^++
\e^{-\frac{\theta}2\,j}\,p_i^{\prime\,+})\,
\P^i_++(\e^{-\frac{\theta}2\,j'}\,p_i^-+
\e^{\frac{\theta}2\,j}\,p_i^{\prime\,-})\,\P^i_-]}
\nonumber\\ &&\times~\e^{\frac\ii2\,(j+j'\,)\,\J}~
\e^{\ii[t+t'-\theta\,(\e^{\frac{\theta}2\,(j+j'\,)}\,
\mbp^+\cdot\mbp^{\prime\,-}-\e^{-\frac{\theta}2\,(j+j'\,)}\,
\mbp^-\cdot\mbp^{\prime\,+})]\,\T} \ . \nonumber\\ &&
\label{TOsymgpprodexpl}\eea

With the same conventions as above, from (\ref{xastarxb}) we may now
compute the star-products $\bullet$ between the coordinate functions
on $\mfn^{\vee\,}$ and again verify the commutation relations of the
algebra $\mfn$,
\bea
x_a\bullet x_a&=&(x_a)^2 \ , \nonumber\\
x_a\bullet x^+&=&x^+\bullet x_a~=~x_a\,x^+ \ ,
\nonumber\\z_1\bullet z_2&=&z_2\bullet z_1
{}~=~z_1\,z_2 \ ,\nonumber\\\overline{z}_1\bullet \overline{z}_2
&=&\overline{z}_2\bullet \overline{z}_1
{}~=~\overline{z}_1\,\overline{z}_2 \ , \nonumber\\
x^-\bullet z_i&=&x^-\,z_i-\mbox{$\frac\ii2$}\,
\theta\,z_i \ , \nonumber\\
z_i\bullet x^-&=&x^-\,z_i+\mbox{$\frac\ii2$}\,
\theta\,z_i \ , \nonumber\\x^-\bullet \overline{z}_i
&=&x^-\,\overline{z}_i+\mbox{$\frac\ii2$}\,
\theta\,\overline{z}_i \ , \nonumber\\\overline{z}_i\bullet x^-
&=&x^-\,\overline{z}_i-\mbox{$\frac\ii2$}\,
\theta\,\overline{z}_i \ , \nonumber\\
z_i\bullet \overline{z}_i&=&z_i\,\overline{z}_i-\ii\theta\,x^+ \ ,
\nonumber\\ \overline{z}_i\bullet z_i &=&z_i\,\overline{z}_i+
\ii\theta\,x^+  \ .
\label{TOsymcoordstarprods}\eea
{}From (\ref{NOproductsBCH},\ref{fstargbidiff}) we find for generic
functions the formula
\bea
f\bullet g&=&\mu\circ\exp\left\{\ii\theta\,x^+\,\left(\e^{-\frac{\ii\theta}2
\,\partial_-}\,
\mdell^\top\otimes\e^{-\frac{\ii\theta}2\,\partial_-}\,
\overline{\mdell}-\e^{\frac{\ii\theta}2\,\partial_-}\,
\overline{\mdell}{}^{\,\top}\otimes\e^{\frac{\ii\theta}2\,\partial_-}\,
\mdell\right)\right.\nonumber
\\ &&\qquad\qquad+\,\overline{z}_i\,\left[\partial^i\otimes
\left(\e^{-\frac{\ii\theta}2\,\partial_-}-1\right)
+\left(\e^{\frac{\ii\theta}2\,\partial_-}-1\right)\otimes
\partial^i\right]\nonumber\\ &&\qquad\qquad+\left.
z_i\,\left[\,\overline{\partial}{}^{\,i}\otimes
\left(\e^{\frac{\ii\theta}2\,\partial_-}-1\right)
+\left(\e^{-\frac{\ii\theta}2\,\partial_-}-1\right)
\otimes\overline{\partial}{}^{\,i}\right]\right\}f\otimes g \ .
\label{TOsymstargen}\eea
To second order in $\theta$ we obtain
\begin{eqnarray}
  \label{eq:symtime:positionspace}\nonumber
  f\bullet g&=&f\,g-\mbox{$\frac{\ii}2$}\,\theta\,\left[
  2x^+\,\left(\,\overline{\mdell}f\cdot\mdell g
  - \mdell f\cdot\overline{\mdell}g\right)\right.\\\nonumber
  &&\qquad\qquad\quad\quad+\left.\overline{\mz}\cdot\left(\mdell f\,\d_- g
  - \d_- f\,\mdell g\right)+ \mz\cdot\left(\d_- f\,\overline{\mdell}g
   - \overline{\mdell}f\,\d_- g\right)\right]\\ \nonumber
  &&-\,\mbox{$\frac1{2}$}\,\theta^2\,\mbox{$\sum\limits_{i=1,2}$}\,
  \left[\left(x^+\right)^2\,\left((\,\overline{\d}{}^{\,i})^2f\,(\d^i)^2g
  +(\d^i)^2f\,(\,\overline{\d}{}^{\,i})^2g
  -2\overline{\d}{}^{\,i}\d^if\,\overline{\d}{}^{\,i}\d^ig
  \right)\right.\\ \nonumber &&\qquad\qquad\quad\quad
  -\,x^+\,\left(\d^if\,\overline{\d}{}^{\,i}\d_- g
  +\overline{\d}{}^{\,i}f\,\d^i\d_- g
  +\overline{\d}{}^{\,i}\d_- f\,\d^ig
  +\d^i\d_- f\,\overline{\d}{}^{\,i}g\right)\\ \nonumber
  &&\qquad\qquad\quad\quad
  +\,x^+\,\overline{z}_i\,\left(\,\overline{\d}{}^{\,i}\d^if\,\d^i\d_- g
  -\overline{\d}{}^{\,i}\d_- f\,(\d^i)^2g
  +\d^i\d_- f\,\overline{\d}{}^{\,i}\d^ig
  -(\d^i)^2f\,\overline{\d}{}^{\,i}\d_- g\right)
  \\ \nonumber &&\qquad\qquad\quad\quad
  +\,x^+\,z_i\,\left(\,\overline{\d}{}^{\,i}\d^if\,\overline{\d}{}^{\,i}\d_- g
  -\d^i\d_- f\,(\,\overline{\d}{}^{\,i})^2g
  +\overline{\d}{}^{\,i}\d_- f\,\overline{\d}{}^i\d^ig
  -(\,\overline{\d}{}^{\,i})^2f\,\d^i\d_- g\right)\\ \nonumber
  &&\qquad\qquad\quad\quad
  +\,\mbox{$\frac12$}\,\overline{z}_i\,z_i\,\left(\,
  \overline{\d}{}^{\,i}\d_- f\,\d^i\d_- g
  +\d^i\d_- f\,\overline{\d}{}^{\,i}\d_- g
  -\d_- ^2f\,\overline{\d}{}^{\,i}\d^ig
  -\overline{\d}{}^{\,i}\d^if\,\d_-^2g\right)
  \\ \nonumber &&\qquad\qquad\quad\quad
  +\,\mbox{$\frac14$}\,\overline{z}_i^{\,2}\,
  \left((\d^i)^2f\,\d_-^2g
  -2\d^i\d_- f\,\d^i\d_- g
  +\d_-^2f\,(\d^i)^2g\right)\\ \nonumber &&\qquad\qquad\quad\quad
  +\,\mbox{$\frac14$}\,z_i^2\,
  \left((\,\overline{\d}{}^{\,i})^2f\,\d_-^2g
  -2\overline{\d}{}^{\,i}\d_- f\,\overline{\d}{}^{\,i}\d_- g
  +\d_-^2f\,(\,\overline{\d}{}^{\,i})^2g\right)
  \\ \nonumber &&\qquad\qquad\quad\quad
  +\Bigl.\mbox{$\frac14$}\,\overline{z}_i\,\left(\d^if\,\d_-^2g
  +\d_-^2f\,\d^ig\right)
  +\mbox{$\frac14$}\,z_i\,\left(\d_-^2f\,\overline{\d}{}^{\,i}g
  +\overline{\d}{}^{\,i}f\,\d_-^2g\right)
  \Bigr]+O\left(\theta^3\right) \ . \\ &&
\end{eqnarray}

\subsection{Weyl Ordering\label{WOP}}

The original Gutt product~\cite{Gutt1} is based on the ``Weyl
ordering'' prescription whereby all monomials  in $U(\mfn)$ are
completely symmetrized over all elements of $\mfn$. On
$\mathcal{N}^\complex$ it is defined by
\beq
\Omega_\star\left(\e^{\ii\mk\cdot\mx}\right)=
\NO\,\e^{\ii k^a\,\X_a}\,\NO:=\e^{\ii k^a\,\X_a} \ .
\label{Weylgpprods}\eeq
While this ordering is usually thought of as the ``canonical''
ordering for the construction of star-products, in our case it turns
out to be drastically more complicated than the other
orderings. Nevertheless, we shall present here its explicit
construction for the sake of completeness and for later comparisons.

It is an extremely arduous task to compute products of the group
elements (\ref{Weylgpprods}) directly from the
Baker-Campbell-Hausdorff formula (\ref{eq:BCH}). Instead, we shall
construct an isomorphism $\mathcal{G}:\overline{U(\mfn)^\complex}\to
\overline{U(\mfn)^\complex}$ which sends the time-ordered product defined by
(\ref{TOgpprods}) into the Weyl-ordered product defined by
(\ref{Weylgpprods}), i.e.
\beq
\mathcal{G}\circ\Omega_*=\Omega_\star \ .
\label{1ststudy}\eeq
Then by defining
$\mathcal{G}_\Omega:=\Omega_*^{-1}\circ\mathcal{G}\circ\Omega^{~}_\star$,
the star-product $\star$ associated with the Weyl ordering
prescription (\ref{Weylgpprods}) may be computed as
\beq
f\star g=\mathcal{G}^{~}_\Omega\bigl(\mathcal{G}_\Omega^{-1}(f)*
\mathcal{G}_\Omega^{-1}(g)\bigr) \ , ~~ f,g\in\CC^\infty(\mfn^{\vee\,}) \ .
\label{WeylTOrel}\eeq
Explicitly, if
\beq
\NOa\,\e^{\ii k^a\,\X_a}\,\NOa=\e^{\ii G^a(\mk)\,\X_a}
\label{1ststudyexpl}\eeq
for some function $\mG=(G^a):\real^6\to\real^6$, then the isomorphism
$\mathcal{G}_\Omega:\CC^\infty(\mfn^{\vee\,})\to\CC^\infty(\mfn^{\vee\,})$ may
be represented as the invertible differential operator
\beq
\mathcal{G}_\Omega=\e^{\ii\mx\cdot[\mG(-\ii\mbf\partial)+\ii\mbf
\partial]} \ .
\label{Gdiffop}\eeq
This relation just reflects the fact that the time-ordered and
Weyl-ordered star-products, although not identical, simply represent
different ordering prescriptions for the same algebra and are
therefore (cohomologically) {\it
  equivalent}. We will elucidate this property more thoroughly in
Section~\ref{WeylSystems}. Thus once the map (\ref{1ststudyexpl}) is
known, the Weyl ordered star-product $\star$ can be computed in terms
of the time-ordered star-product $*$ of Section~\ref{TOP}.

The functions $G^a(\mk)$ appearing in (\ref{1ststudyexpl}) are readily
calculable through the Baker-Campbell-Hausdorff formula. It is clear
from (\ref{TOgpprods}) that the coefficient of the time translation
generator $\J\in\mfn$ is simply
\beq
G^j(j,t,\mbp^\pm)=j \ .
\label{Gj}\eeq
{}From (\ref{eq:BCH}) it is also clear that the only terms proportional
to $\P^i_+$ come from commutators of the form
$[\J,[\dots,[\J,\P^i_+]\,]\dots]$, and gathering all terms we find
\bea
\mbox{$\sum\limits_{i=1,2}$}\,
G^{p_i^+}(j,t,\mbp^\pm)~\P^i_+&=&-\ii\sum_{n=0}^\infty
\frac{B_n}{n!}~\bigl[~\underbrace{\ii j\,\J\,,\,\bigl[\dots\,,\,
\bigl[\ii j\J}_n\,,\,\ii p_i^+\,\P^i_+\,\bigr]\,\bigr]
\dots\bigr]\nonumber\\ &=&
p_i^+\,\sum_{n=0}^\infty\frac{B_n}{n!}\,
(-\theta\,j)^n~\P^i_+ \ .
\label{GomzBn}\eea
Since $B_0=1$, $B_1=-\frac12$ and $B_{2k+1}=0~~\forall k\geq1$, from
(\ref{eq:BCH:K}) we thereby find
\beq
G^{\mbp^+}(j,t,\mbp^\pm)=\frac{\mbp^+}
{\phi_\theta(j)}
\label{Gomz}\eeq
where we have introduced the function
\beq
\phi_\theta(j)=\frac{1-\e^{-\theta\,j}}{\theta\,j}
\label{phithetadef}\eeq
obeying the identities
\beq
\phi_\theta(j)~\e^{\theta\,j}=\phi_{-\theta}(j) \ , \quad
\phi_\theta(j)\,\phi_{-\theta}(j)=-\frac2{(\theta\,j)^2}\,\bigl(
1-\cos(\theta\,j)\bigr) \ .
\label{phithetaids}\eeq
In a completely analogous way one finds the coefficient of the
$\P^i_-$ term to be given by
\beq
G^{\mbp^-}(j,t,\mbp^\pm)=\frac{\mbp^-}
{{\phi_{-\theta}(j)}} \ .
\label{Gmz}\eeq
Finally, the non-vanishing contributions to the central
element $\T\in\mfn$ are given by
\bea
G^t(j,t,\mbp^\pm)~\T&=&
t~\T-\ii\sum_{n=1}^\infty\frac{B_{n+1}}{n!}\,
\left(\bigl[\ii p_i^+\,\P^i_+\,,\,\bigl[~
\underbrace{\ii j\,\J\,,\,\dots\bigl[\ii j\,\J}_n\,,\,\ii p_i^-\,
\P^i_-\,\bigr]\dots\bigr]\,\bigr]\right.\nonumber\\ &&\qquad\qquad\qquad
+\left.\bigl[\ii p_i^-\,\P^i_-\,,\,\bigl[~
\underbrace{\ii j\,\J\,,\,\dots\bigl[\ii j\,\J}_n\,,\,
\ii p_i^+\,\P^i_+\,\bigr]\dots\bigr]\,\bigr]\right)
\nonumber\\ &=&t~\T+4\theta\,\mbp^+\cdot\mbp^-\,\sum_{n=1}^\infty\frac{B_{n+1}}
{n!}\,(-\theta\,j)^n~\T \ .
\label{GtBn}\eea
By differentiating (\ref{GomzBn}) and (\ref{phithetadef}) with respect
to $s=-\theta\,j$ we arrive finally at
\beq
G^t(j,t,\mbp^\pm)=t+4\theta\,\mbp^+\cdot\mbp^-\,
\gamma_\theta(j)
\label{Gt}\eeq
where we have introduced the function
\beq
\gamma_\theta(j)=\frac12+\frac{(1+\theta\,j)~\e^{-\theta\,j}-1}
{\left(\e^{-\theta\,j}-1\right)^2} \ .
\label{gammathetadef}\eeq

{}From (\ref{Gdiffop}) we may now write down the explicit form of the
differential operator implementing the equivalence between the
star-products $*$ and $\star$ as
\bea
\mathcal{G}_\Omega&=&\exp\left[-2\ii\theta\,x^+\,\overline{\mdell}
\cdot\mdell\left(1+\frac{2(1-\ii\theta\,\partial_-)~
\e^{\ii\theta\,\partial_-}-1}{\left(\e^{\ii\theta\,\partial_-}-1
\right)^2}\right)\right.\nonumber\\ &&\qquad\quad+\left.
\overline{\mz}\cdot\mdell\left(\frac{\ii\theta\,\partial_-}
{\e^{\ii\theta\,\partial_-}-1}-1\right)-\mz\cdot\overline{\mdell}
\left(\frac{\ii\theta\,\partial_-}
{\e^{-\ii\theta\,\partial_-}-1}+1\right)\right] \ .
\label{Gdiffopexpl}\eea
{}From (\ref{TOgpprodexpl}) and (\ref{1ststudyexpl}) we may readily
compute the products of Weyl symbols with the result
\bea
&&\NO~~\NO\,\e^{\ii k^a\,\X_a}\,\NO\cdot\NO\,\e^{\ii k^{\prime\,a}\,\X_a}
\,\NO~~\NO \nonumber\\ && {~~~~}_{~~}^{~~}
\nn\\ &&~=~\exp\ii\left\{\frac{\phi_\theta(j)\,p_i^++
\e^{-\theta\,j}\,\phi_\theta(j'\,)\,p_i^{\,\prime\,+}}
{\phi_\theta(j+j'\,)}~\P^i_+
+\frac{{\phi_{-\theta}(j)}\,p_i^-+
\e^{\theta\,j}\,{\phi_{-\theta}(j'\,)}\,p_i^{\prime\,-}}
{{\phi_{-\theta}(j+j'\,)}}~\P^i_-\right.\nonumber\\
&&\quad\qquad\qquad+\,(j+j'\,)~\J+\left[t+t'+\theta\,
\bigl(\,{\phi_{-\theta}(j)\,
\phi_{-\theta}(j'\,)}\,\mbp^+\cdot\mbp^{\prime\,-}-{\phi_{\theta}(j)\,
\phi_{\theta}(j'\,)}\,\mbp^-\cdot\mbp^{\prime\,+}\bigr)\right.
\nonumber\\ &&\quad\qquad\qquad-\,4\theta\,\left(\gamma_\theta(j+j'\,)
\,\bigl({\phi_{-\theta}(j)}\,
\mbp^++\e^{\theta\,j}\,{\phi_{-\theta}(j'\,)}\,\mbp^{\prime\,+}\,\bigr)
\cdot\bigl(\,\phi_\theta(j)\,
\mbp^-+\e^{-\theta\,j}\,\phi_\theta(j'\,)\,\mbp^{\prime\,-}\,\bigr)
\right.\nonumber\\ &&\quad\qquad\qquad\quad\qquad-\biggl.\left.\left.
\gamma_\theta(j)\,\phi_\theta(j)\,\phi_{-\theta}(j)\,\mbp^+\cdot\mbp^--
\gamma_\theta(j'\,)\,\phi_\theta(j'\,)\,\phi_{-\theta}(j'\,)
\,\mbp^{\prime\,+}
\cdot\mbp^{\prime\,-}\right)\right]~\T\biggr\} \ . \nonumber\\ &&
\label{Weylgpprodexpl}\eea
{}From (\ref{xastarxb}) we may now compute the star-products $\star$
between the coordinate functions on $\mfn^{\vee\,}$ to be
\bea
x_a\star x_a&=&(x_a)^2 \ , \nonumber\\
x_a\star x^+&=&x^+\star x_a~=~x_a\,x^+ \ ,
\nonumber\\z_1\star z_2&=&z_2\star z_1
{}~=~z_1\,z_2 \ ,\nonumber\\\overline{z}_1\star \overline{z}_2
&=&\overline{z}_2\star \overline{z}_1
{}~=~\overline{z}_1\,\overline{z}_2 \ , \nonumber\\
x^-\star z_i&=&x^-\,z_i-\mbox{$\frac\ii2$}\,
\theta\,z_i \ , \nonumber\\
z_i\star x^-&=&x^-\,z_i+\mbox{$\frac\ii2$}\,
\theta\,z_i \ , \nonumber\\x^-\star \overline{z}_i
&=&x^-\,\overline{z}_i+\mbox{$\frac\ii2$}\,
\theta\,\overline{z}_i \ , \nonumber\\\overline{z}_i\star x^-
&=&x^-\,\overline{z}_i-\mbox{$\frac\ii2$}\,
\theta\,\overline{z}_i \ , \nonumber\\
z_i\star \overline{z}_i&=&z_i\,\overline{z}_i-\ii\theta\,x^+ \ ,
\nonumber\\ \overline{z}_i\star z_i &=&z_i\,\overline{z}_i+
\ii\theta\,x^+  \ .
\label{Weylsymcoordstarprods}\eea
These products are identical to those of the symmetric time ordering
prescription (\ref{TOsymcoordstarprods}). After some computation, from
(\ref{NOproductsBCH},\ref{fstargbidiff}) we find for generic functions
$f,g\in\CC^\infty(\mfn^{\vee\,})$ the formula
\bea
f\star g&=&\mu\circ\exp\left\{\theta\,x^+\,\left[~~
\frac{1\otimes1+\bigl(\ii
\theta\,(\partial_-\otimes1+1\otimes\partial_-)-1\otimes1
\bigr)~\e^{\ii\theta\,\partial_-}\otimes\e^{\ii\theta\,\partial_-}}
{\left(\e^{\ii\theta\,\partial_-}\otimes\e^{\ii\theta\,\partial_-}-1\otimes1
\right)^2}\right.\right.\nonumber\\ &&\times\,
\left(\frac{4\mdell^\top\left(\e^{-\ii\theta\,\partial_-}-1
\right)}{\theta\,\partial_-}\otimes
\frac{\overline{\mdell}\left(\e^{-\ii\theta\,\partial_-}-1
\right)}{\theta\,\partial_-}-\frac{3\overline{\mdell}{}^{\,\top}
\left(\e^{\ii\theta\,\partial_-}-1
\right)}{\theta\,\partial_-}\otimes
\frac{\mdell\left(\e^{\ii\theta\,\partial_-}-1
\right)}{\theta\,\partial_-}\right.\nonumber\\ &&+\left.
\frac{4\overline{\mdell}\cdot\mdell\sin^2\left(\frac\theta2
\,\partial_-\right)}{\theta^2\,\partial_-^2}\otimes1-1\otimes
\frac{4\overline{\mdell}\cdot\mdell\sin^2\left(\frac\theta2
\,\partial_-\right)}{\theta^2\,\partial_-^2}\right)
\nonumber\\ &&+\,
\frac{4\ii\overline{\mdell}\cdot\mdell}{\theta\,\partial_-}
\left(\frac{\ii\sin^2\left(\frac\theta2
\,\partial_-\right)}{\theta\,\partial_-\left(
\e^{\ii\theta\,\partial_-}-1\right)}-1\right)\otimes1+1\otimes
\frac{4\ii\overline{\mdell}\cdot\mdell}{\theta\,\partial_-}
\left(\frac{\ii\sin^2\left(\frac\theta2
\,\partial_-\right)}{\theta\,\partial_-\left(
\e^{\ii\theta\,\partial_-}-1\right)}-1\right)\nonumber\\ &&
+\left.\frac{3\overline{\mdell}{}^{\,\top}\left(\e^{\ii\theta\,\partial_-}-1
\right)}{\theta\,\partial_-}\otimes\frac{
\mdell\left(\e^{\ii\theta\,\partial_-}-1
\right)}{\theta\,\partial_-}+\frac{\mdell^\top
\left(\e^{-\ii\theta\,\partial_-}-1
\right)}{\theta\,\partial_-}\otimes\frac{\overline{\mdell}
\left(\e^{-\ii\theta\,\partial_-}-1\right)}{\theta\,\partial_-}~~\right]
\nonumber\\&&
+\,\frac{\overline{z}_i}{1\otimes\e^{-\ii\theta\,
\partial_-}-\e^{\ii\theta\,\partial_-}\otimes1}\,\left[~~
\frac{\d^i}{\partial_-}\,\left(1-\e^{\ii\theta\,\partial_-}
\right)\otimes\partial_--\partial_-\otimes\frac{\d^i}
{\partial_-}\,\left(1-\e^{-\ii\theta\,\partial_-}\right)\right.\nonumber\\
&& +\Biggl.\,1\otimes\d^i~
\e^{-\ii\theta\,\partial_-}-\d^i~\e^{\ii\theta\,\partial_-}
\otimes1-1\otimes2\d^i~~\Biggr]\nonumber\\ &&
+\,\frac{z_i}{1\otimes\e^{\ii\theta\,
\partial_-}-\e^{-\ii\theta\,\partial_-}\otimes1}\,\left[~~
\frac{\overline{\d}{}^{\,i}}{\partial_-}\,\left(1-\e^{-\ii\theta\,\partial_-}
\right)\otimes\partial_--\partial_-\otimes\frac{\overline{\d}{}^{\,i}}
{\partial_-}\,\left(1-\e^{\ii\theta\,\partial_-}\right)\right.\nonumber\\
&& +\left.\Biggl.1\otimes\overline{\d}{}^{\,i}~
\e^{\ii\theta\,\partial_-}-\overline{\d}{}^{\,i}~\e^{-\ii\theta\,\partial_-}
\otimes1-1\otimes2\overline{\d}{}^{\,i}~~\Biggr]\right\}f\otimes g \ .
\label{Weylstargen}\eea
To second order in the deformation parameter $\theta$ we obtain
\begin{eqnarray}
  \label{eq:weyl:positionspace}
  f\star g&=&f\,g-\mbox{$\frac{\ii}2$}\,\theta\,\left[
  2x^+\,\left(\,\overline{\mdell}f\cdot\mdell g
  - \mdell f\cdot\overline{\mdell}g\right)\right.\\\nonumber
  &&\qquad\qquad\qquad+\left.\overline{\mz}\cdot\left(\mdell f\,\d_- g
  - \d_- f\,\mdell g\right)+ \mz\cdot\left(\d_- f\,\overline{\mdell}g
   - \overline{\mdell}f\,\d_- g\right)\right]\\ \nonumber
  &&-\,\mbox{$\frac1{2}$}\,\theta^2\,\mbox{$\sum\limits_{i=1,2}$}\,\left[
  \left(x^+\right)^2\,
  \left((\,\overline{\d}{}^{\,i})^2f\,(\d^i)^2g
  +(\d^i)^2f\,(\,\overline{\d}{}^{\,i})^2g
  -2\overline{\d}{}^{\,i}\d^if\,\overline{\d}{}^{\,i}\d^ig
  \right)\right.\\ \nonumber &&\qquad\qquad\quad\quad
  -\,\mbox{$\frac13$}\,x^+\,\left(\d^if\,\overline{\d}{}^{\,i}\d_- g
  +\overline{\d}{}^{\,i}f\,\d^i\d_- g
  +\overline{\d}{}^{\,i}\d_- f\,\d^ig
  +\d^i\d_- f\,\overline{\d}{}^{\,i}g\right.\\ \nonumber
  &&\qquad\qquad\qquad\qquad\qquad\quad\quad
  -\left.2\partial_-f\,\overline{\d}{}^{\,i}
  \d^ig-2\overline{\d}{}^{\,i}\partial^if\,
  \partial_-g\right)\\\nonumber&&\qquad\qquad\quad\quad
  +\,x^+\,\overline{z}_i\,\left(\,\overline{\d}{}^{\,i}
  \d^if\,\d^i\d_- g
  -\overline{\d}{}^{\,i}\d_- f\,(\d^i)^2g
  +\d^i\d_- f\,\overline{\d}{}^{\,i}\d^ig
  -(\d^i)^2f\,\overline{\d}{}^{\,i}\d_- g\right)
  \\ \nonumber &&\qquad\qquad\quad\quad
  +\,x^+\,z_i\,\left(\,\overline{\d}{}^{\,i}\d^if\,\overline{\d}{}^{\,i}\d_- g
  -\d^i\d_- f\,(\,\overline{\d}{}^{\,i})^2g
  +\overline{\d}{}^{\,i}\d_- f\,\overline{\d}{}^{\,i}\d^ig
  -(\,\overline{\d}{}^{\,i})^2f\,\d^i\d_- g\right)\\ \nonumber
  &&\qquad\qquad\quad\quad
  +\,\mbox{$\frac12$}\,\overline{z}_i\,z_i\,\left(\,
  \overline{\d}{}^{\,i}\d_- f\,\d^i\d_- g
  +\d^i\d_- f\,\overline{\d}{}^{\,i}\d_- g
  -\d_- ^2f\,\overline{\d}{}^{\,i}\d^ig
  -\overline{\d}{}^{\,i}\d^if\,\d_-^2g\right)
  \\ \nonumber &&\qquad\qquad\quad\quad
  +\,\mbox{$\frac14$}\,\overline{z}_i^{\,2}\,
  \left((\d^i)^2f\,\d_-^2g
  -2\d^i\d_- f\,\d^i\d_- g
  +\d_-^2f\,(\d^i)^2g\right)\\ \nonumber &&\qquad\qquad\quad\quad
  +\,\mbox{$\frac14$}\,z_i^2\,
  \left((\,\overline{\d}{}^{\,i})^2f\,\d_-^2g
  -2\overline{\d}{}^{\,i}\d- f\,\overline{\d}{}^{\,i}\d_- g
  +\d_-^2f\,(\,\overline{\d}{}^{\,i})^2g\right)
  \\ \nonumber &&\qquad\qquad\quad\quad
  +\,\mbox{$\frac16$}\,\overline{z}_i\,\left(\d^if\,\d_-^2g
  +\d_-^2f\,\d^ig-\partial_-f\,\partial^i
  \partial_-g-\partial^i\partial_-f\,\partial_-g\right)
  \\&&\qquad\qquad\quad\quad
  +\Bigl.\mbox{$\frac16$}\,z_i\,\left(\d_-^2f\,\overline{\d}{}^{\,i}g
+\overline{\d}{}^{\,i}f\,\d_-^2g-\partial_-f\,
\overline{\partial}{}^{\,i}\d_-g-
  \overline{\d}{}^{\,i}\d_-f\,\d_-g\right)
  \Bigr]+O\left(\theta^3\right) \ . \nn\\ &&
\end{eqnarray}

Although extremely cumbersome in form, the Weyl-ordered product has
several desirable features over the simpler time-ordered products. For
instance, the Schwartz subspace of $\CC^\infty(\mfn^{\vee\,})$ is closed
under the Weyl ordered product, whereas the other products are only
formal in this regard and do not define strict deformation
quantizations. It is also hermitean owing to the property
\beq
\overline{f\star g}=\overline{g}\star\overline{f} \ .
\label{Weylstarherm}\eeq
Moreover, while the $\mfn$-covariance condition (\ref{xyPoisson})
holds for all of our star-products, the Weyl product is in fact
$\mfn$-invariant, because for any $x\in\complex_{(1)}(\mfn^{\vee\,})$ one
has the stronger compatibility condition
\beq
[x,f]_\star=\ii\theta\,\Theta(x,f) ~~~~ \forall f\in\CC^\infty(\mfn^{\vee\,})
\label{strongcompcond}\eeq
with the action of the Lie algebra $\mfn$. In the next
section we shall see that the Weyl-ordered star-product is, in a
certain sense, the generator of all other star-products making it the
``universal'' product for the quantization of the spacetime $\NW_6$.

\newsection{Weyl Systems\label{WeylSystems}}

In this section we will use the notion of a generalized Weyl system
introduced in~\cite{ALZ1} to describe some more formal aspects of the
star-products that we have constructed and to analyse the interplay
between them. This generalizes the standard Weyl systems~\cite{Sz1}
which may be used to provide a purely operator theoretic
characterization of the Moyal product, associated to the (untwisted)
Heisenberg algebra. In that case, it can be regarded as a projective
representation of the translation group in an even-dimensional real
vector space. However, for the twisted Heisenberg algebra such a
representation is not possible, since by definition the appropriate
arena should be a central extension of the non-abelian subgroup
$\mathcal S$ of the full euclidean group ${\rm ISO}(4)$. This requires a
generalization of the standard notion which we will now describe and
use it to obtain a very useful characterization of the noncommutative
geometry induced by the algebra $\mfn$.

Let $\mathbb{V}$ be a five-dimensional real vector space. In a
suitable (canonical) basis, vectors
$\mk\in\mathbb{V}\cong\real\times\complex^2$ will be denoted (with
respect to a chosen complex structure) as
\beq
\mk=\begin{pmatrix}j\\\mbp^+\,\\\mbp^-\end{pmatrix}
\label{Svectors}\eeq
with $j\in\real$ and $\mbp^\pm=\overline{\mbp^\mp}\in\complex^2$. As
the notation suggests, we
regard $\mathbb{V}$ as the ``momentum space'' of the dual
$\mfn^{\vee\,}$. Note that we do not explicitly incorporate the component
corresponding to the central element $\T$, as it will instead appear
through the appropriate projective representation that we will construct,
similarly to the Moyal case. As an abelian group,
$\mathbb{V}\cong\real^5$ with the usual addition $+$ and identity
$\mbf0$. Corresponding to a deformation parameter
$\theta\in\real$, we deform this abelian Lie group structure to a
generically non-abelian one. The deformed composition law is denoted
$\comp$. It is associative and in general will depend on
$\theta$. The identity element with respect to $\comp$ is still
defined to be $\mbf0$, and the inverse of any element $\mk\in\mathbb{V}$
is denoted $\underline{\mk}$, so that
\beq
\mk\comp\underline{\mk}=\underline{\mk}\comp\mk=\mbf0 \ .
\label{compinverse}\eeq
Being a deformation of the underlying abelian group structure on
$\mathbb{V}$ means that the composition of any two vectors
$\mk,\mq\in\mathbb{V}$ has a formal small $\theta$ expansion of the
form
\beq
\mk\comp\mq=\mk+\mq+O(\theta) \ ,
\label{compsmalltheta}\eeq
from which it follows that
\beq
\underline{\mk}=-\mk+O(\theta) \ .
\label{compinvsmalltheta}\eeq
In other words, rather than introducing star-products that deform the
pointwise multiplication of functions on $\mfn^{\vee\,}$, we now deform the
``momentum space'' of $\mfn^{\vee\,}$ to a non-abelian Lie group.
We will see below that the five-dimensional group
$(\mathbb{V},\comp)$ is isomorphic to the original subgroup
$\mathcal{S}\subset{\rm ISO}(4)$, and that the two notions of
quantization are in fact the same.

Given such a group, we now define a (generalized) Weyl system for the
algebra $\mfn$ as a quadruple
$(\mathbb{V},\comp,\weyl,\omega)$, where the map
\beq
\weyl\,:\,\mathbb{V}~\longrightarrow~\overline{U(\mfn)^\complex}
\label{genweylmapdef}\eeq
is a projective representation of the group
$(\mathbb{V},\comp)$ with projective phase
$\omega:\mathbb{V}\times\mathbb{V}\to\complex$. This means that for
every pair of elements $\mk,\mq\in\mathbb{V}$ one has the composition
rule
\beq
\weyl(\mk)\cdot\weyl(\mq)=\e^{\frac\ii2\,\omega(\mk,\mq)\,\T}~\cdot~
\weyl(\mk\comp\mq)
\label{Weylcomprule}\eeq
in the completed, complexified universal enveloping algebra of
$\mfn$. The associativity of $\comp$ and the relation
(\ref{Weylcomprule}) imply that the subalgebra
$\weyl(\mathbb{V})\subset\overline{U(\mfn)^\complex}$ is
associative if and only if
\beq
\omega(\mk\comp\mbf p,\mq)=\omega(\mk,\mbf p\comp\mq)+\omega(\mbf p,\mq)-
\omega(\mk,\mbf p)
\label{cocyclecond}\eeq
for all vectors $\mk,\mq,\mbf p\in\mbbV$. This condition means that
$\omega$ defines a one-cocycle in the group cohomology
of $(\mathbb{V},\comp)$. It is automatically satisfied if
$\omega$ is a bilinear form with respect to $\comp$. We will in
addition require that $\omega(\mk,\mq)=O(\theta)~~\forall
\mk,\mq\in\mathbb{V}$ for consistency with
(\ref{compsmalltheta}). The identity element of $\weyl(\mathbb{V})$
is $\weyl(\mbf0)$ while the inverse of $\weyl(\mk)$ is given by
\beq
\weyl(\mk)^{-1}=\weyl(\,\underline{\mk}\,) \ .
\label{weylinverse}\eeq
The standard Weyl system on $\real^{2n}$ takes $\comp$ to be ordinary
addition and $\omega$ to be the Darboux symplectic two-form, so that
$\weyl(\real^{2n})$ is a projective representation of the translation
group, as is appropriate to the Moyal product.

Given a Weyl system defined as above, we can now introduce
another isomorphism
\beq
\Pi\,:\,\CC^\infty\left(\real^5\right)~\longrightarrow~
\weyl\bigl(\mathbb{V}\bigr)
\label{Omegaquantmap}\eeq
defined by the symbol
\beq
\Pi(f):=\int\limits_{\real^5}\frac{\dd\mk}{(2\pi)^5}~
\tilde{f}(\mk)~\weyl(\mk)
\label{Omegafdef}\eeq
where as before $\tilde f$ denotes the Fourier transform of
$f\in\CC^\infty(\real^5)$. This definition implies that
\beq
\Pi\bigl(\e^{\ii\mk\cdot\mx}\bigr)=\weyl(\mk) \ ,
\label{weylmkDelta}\eeq
and that we may introduce a $*$-involution $\dag$ on both algebras
$\CC^\infty(\real^5)$ and $\weyl(\mbbV)$ by the formula
\beq
\Pi\bigl(f^\dag\bigr)=\Pi\bigl(f\bigr)^\dag:=
\int\limits_{\real^5}\frac{\dd\mk}{(2\pi)^5}~\overline{
\tilde f(\,\underline{\mk}\,)}~\weyl(\mk) \ .
\label{involdef}\eeq
The compatibility condition
\beq
\bigl(\Pi(f)\cdot\Pi(g)\bigr)^\dag=\Pi(g)^\dag\cdot
\Pi(f)^\dag
\label{compconddag}\eeq
with the product in $\overline{U(\mfn)^\complex}$ imposes further
constraints on the group composition law $\comp$ and cocycle
$\omega$~\cite{ALZ1}. From (\ref{Weylcomprule}) we may thereby define
a $\dag$-hermitean star-product of $f,g\in\CC^\infty(\real^5)$ by the
formula
\beq
f\star g:=\Pi^{-1}\bigl(\Pi(f)\cdot\Pi(g)\bigr)
=\int\limits_{\real^5}\frac{\dd\mk}{(2\pi)^5}~
\int\limits_{\real^5}\frac{\dd\mq}{(2\pi)^5}~\tilde f(\mk)\,
\tilde g(\mq)~\e^{\frac\ii2\,\omega(\mk,\mq)}~
\Pi^{-1}\circ\weyl(\mk\comp\mq) \ ,
\label{fstargweyl}\eeq
and in this way we have constructed a quantization of the
algebra $\mfn$ solely from the formal notion of a Weyl system.
The associativity of $\star$ follows from associativity of
$\comp$. We may also rewrite the star-product (\ref{fstargweyl}) in
terms of a bi-differential operator as
\beq
f\star g=f~\e^{\frac\ii2\,\omega(\,-\ii\overleftarrow{\mbf\d}\,,\,
-\ii\overrightarrow{\mbf\d}\,)+\ii\mx\cdot(-\ii\overleftarrow{\mbf\d}\,\comp
-\ii\overrightarrow{\mbf\d}+\ii\overleftarrow{\mbf\d}+\ii
\overrightarrow{\mbf\d}\,)}~g \ .
\label{bidiffweyl}\eeq

This deformation is completely characterized in terms of the
new algebraic structure and its projective representation provided by
the Weyl system. It is straightforward to show
that the Lie algebra of $(\mathbb{V},\comp)$ coincides precisely with
the original subalgebra $\mfs\subset{\rm iso}(4)$, while the cocycle $\omega$
generates the central extension of $\mfs$ to $\mfn$ in the
usual way. From (\ref{fstargweyl}) one may compute the
star-products of coordinate functions on $\real^5$ as
\beq
x_a\star x_b=x_a\,x_b-\ii\mx\cdot\left.\frac\d{\d k^a}\frac\d
{\d q^b}(\mk\comp\mq)\right|_{\mk=\mq=\mbf0}-\left.\frac\ii2\,
\frac\d{\d k^a}\frac\d
{\d q^b}\omega(\mk,\mq)\right|_{\mk=\mq=\mbf0} \ .
\label{xastarxbweyl}\eeq
The corresponding star-commutator may thereby be written as
\beq
[x_a,x_b]_\star=\ii\theta\,C_{ab}^{~~c}\,x_c+\ii\theta\,\xi_{ab} \ ,
\label{xaxbstarcommxi}\eeq
where the relation
\beq
\theta\,C_{ab}^{~~c}=-\left.\left(\frac\d{\d k^a}\frac\d
{\d q^b}-\frac\d{\d k^b}\frac\d
{\d q^a}\right)(\mk\comp\mq)^c\right|_{\mk=\mq=\mbf0}
\label{Cabccomp}\eeq
gives the structure constants of the Lie algebra defined by the Lie
group $(\mathbb{V},\comp)$, while the cocycle term
\beq
\theta\,\xi_{ab}=-\frac12\,\left.\left(\frac\d{\d k^a}\frac\d
{\d q^b}-\frac\d{\d k^b}\frac\d
{\d q^a}\right)\omega(\mk,\mq)\right|_{\mk=\mq=\mbf0}
\label{cocycleterm}\eeq
gives the usual form of a central extension of this Lie
algebra. Demanding that this yield a deformation quantization of the
Kirillov-Kostant Poisson structure on $\mfn^{\vee\,}$ requires that
$C_{ab}^{~~c}$ coincide with the structure constants of the subalgebra
$\mfs\subset{\rm iso}(4)$ of $\mfn$, and also that
$\xi_{\mbp^-,\mbp^+}=-\xi_{\mbp^+,\mbp^-}=2t$ be the only
non-vanishing components of the central extension.

It is thus possible to define a broad class of deformation quantizations of
$\mfn^{\vee\,}$ solely in terms of an abstract Weyl system
$(\mbbV,\comp,\weyl,\omega)$, without explicit realization of the
operators $\weyl(\mk)$. In the remainder of this section we will
set $\Pi=\Omega$ above and describe the Weyl systems
underpinning the various products that we constructed previously. This
entails identifying the appropriate maps (\ref{genweylmapdef}), which
enables the calculation of the projective representations
(\ref{Weylcomprule}) and hence explicit realizations of the group
composition laws $\comp$ in the various instances. This unveils a
purely algebraic description of the star-products which will be
particularly useful for our later constructions, and enables one to
make the equivalences between these products explicit.

\subsection{Time Ordering\label{TOPGWS}}

Setting $t=t'=0$ in (\ref{TOgpprodexpl}), we find the ``time-ordered''
non-abelian group composition law $\compa$ for any two elements of the
form (\ref{Svectors}) to be given by
\beq
\mk\compa\mk'=\begin{pmatrix}j+j'\,\\\mbp^++\e^{-\theta\,j}\,
\mbp^{\prime\,+}\,\\\mbp^-+\e^{\theta\,j}\,\mbp^{\prime\,-}
\,\end{pmatrix} \ .
\label{TOcomplaw}\eeq
{}From (\ref{TOcomplaw}) it is straightforward to compute the inverse
$\underline{\mk}$ of a group element (\ref{Svectors}), satisfying
(\ref{compinverse}), to be
\beq
\underline{\mk}=-\begin{pmatrix}j\\\e^{\theta\,j}\,\mbp^+\\
\e^{-\theta\,j}\,\mbp^-\,\end{pmatrix} \ .
\label{TOinverse}\eeq
The group cocycle is given by
\beq
\omega_*(\mk,\mk'\,)=2\ii\theta\,\left(\e^{\theta\,j}\,
\mbp^+\cdot\mbp^{\prime\,-}-\e^{-\theta\,j}\,
\mbp^-\cdot\mbp^{\prime\,+}\right)
\label{TOcocycle}\eeq
and it defines the canonical symplectic structure on the $j={\rm
  constant}$ subspaces $\complex^2\subset\mathbb{V}$. Note that in
  this representation, the central coordinate function $x^+$ is not
  written explicitly and is simply understood as the unit element of
  $\complex(\real^5)$, as is conventional in the case of the Moyal
  product. For $\mk\in\mathbb{V}$ and $\X_a\in\mfs$ the projective
  representation (\ref{Weylcomprule}) is generated by the time-ordered
  group elements
\beq
\weyl_*(\mk)=\NOa\,\e^{\ii k^a\,\X_a}\,\NOa
\label{TOweylop}\eeq
defined in (\ref{eq:time:defn}).

\subsection{Symmetric Time Ordering\label{TSOPGWS}}

In a completely analogous manner, inspection of
(\ref{TOsymgpprodexpl}) reveals the ``symmetric time-ordered''
non-abelian group composition law $\compb$ defined by
\beq
\mk\compb\mk'=\begin{pmatrix}j+j'\,\\\e^{\frac{\theta}2\,j'}\,\mbp^++
\e^{-\frac{\theta}2\,j}\,\mbp^{\prime\,+}\,\\\e^{-\frac{\theta}2\,j'}\,
\mbp^-+\e^{\frac{\theta}2\,j}\,\mbp^{\prime\,-}\,\end{pmatrix} \ ,
\label{TOsymcomplaw}\eeq
for which the inverse $\underline{\mk}$ of a group element
(\ref{Svectors}) is simply given by
\beq
\underline{\mk}=-\mk \ .
\label{TOsyminverse}\eeq
The group cocycle is
\beq
\omega_\bullet(\mk,\mk'\,)=2\ii\theta\,\left(\e^{\frac{\theta}2\,
(j+j'\,)}\,\mbp^+\cdot\mbp^{\prime\,-}-\e^{-\frac{\theta}2\,
(j+j'\,)}\,\mbp^-\cdot\mbp^{\prime\,+}\right)
\label{TOsymcocycle}\eeq
and it again induces the canonical symplectic structure on
$\complex^2\subset\mbbV$. The corresponding projective representation
of $(\mbbV,\compb)$ is generated by the symmetric time-ordered group
elements
\beq
\weyl_\bullet(\mk)=\NOb\,\e^{\ii k^a\,\X_a}\,\NOb
\label{TOsymweylop}\eeq
defined in (\ref{TOsymgpprods}).

\subsection{Weyl Ordering\label{WOPGWS}}

Finally, we construct the Weyl system
$(\mbbV,\compc,\weyl_\star,\omega_\star)$ associated with the
Weyl-ordered star-product of Section~\ref{WOP}. Starting from
(\ref{Weylgpprodexpl}) we introduce the non-abelian group composition
law $\compc$ by
\beq
\mk\compc\mk'=\begin{pmatrix}j+j'\,\\[2mm]
\frac{\phi_\theta(j)\,\mbp^++
\e^{-\theta\,j}\,\phi_\theta(j'\,)\,\mbp^{\prime\,+}}
{\phi_\theta(j+j'\,)}\\[3mm]
\frac{\phi_{-\theta}(j)\,\mbp^-+
\e^{\theta\,j}\,\phi_{-\theta}(j'\,)\,\mbp^{\prime\,-}}
{\phi_{-\theta}(j+j'\,)}\end{pmatrix} \ ,
\label{Weylcomplaw}\eeq
from which we may again straightforwardly compute the inverse
$\underline{\mk}$ of a group element (\ref{Svectors}) simply as
\beq
\underline{\mk}=-\mk \ .
\label{Weylinverse}\eeq
When combined with the definition (\ref{involdef}), one has
$f^\dag=\overline{f}~~\forall f\in\CC^\infty(\real^5)$ and this
explains the hermitean property (\ref{Weylstarherm}) of the
Weyl-ordered star-product $\star$. This is also true of the product
$\bullet$, whereas $*$ is only hermitean with respect to the modified
involution $\dag$ defined by (\ref{involdef})
and~(\ref{TOinverse}). The group cocycle is given by
\bea
\omega_\star(\mk,\mk'\,)&=&-2\ii\theta\,\Bigl(
\phi_{-\theta}(j)\,\phi_{-\theta}(j'\,)\,
\mbp^+\cdot\mbp^{\prime\,-}-\phi_{\theta}(j)\,\phi_{\theta}(j'\,)\,
\mbp^-\cdot\mbp^{\prime\,+}\Bigr.\nonumber\\ &&\qquad
-\,\gamma_\theta(j+j'\,)\,\bigl(\phi_\theta(j)\,
\mbp^++\e^{-\theta\,j}\,\phi_\theta(j'\,)\,\mbp^{\prime\,+}\bigr)
\cdot\bigl(\phi_{-\theta}(j)\,
\mbp^-+\e^{\theta\,j}\,\phi_{-\theta}(j'\,)\,
\mbp^{\prime\,-}\bigr)\nonumber\\ &&
\qquad+\Bigl.
\gamma_\theta(j)\,\phi_\theta(j)\,\phi_{-\theta}(j)\,\mbp^+\cdot\mbp^-+
\gamma_\theta(j'\,)\,\phi_\theta(j'\,)\,\phi_{-\theta}(j'\,)\,
\mbp^{\prime\,+}\cdot\mbp^{\prime\,-}
\Bigr) \ .
\label{Weylcocycle}\eea
In contrast to the other cocycles, this does {\it not} induce any
symplectic structure, at least not in the manner described
earlier. The corresponding projective representation
(\ref{Weylcomprule}) is generated by the completely symmetrized group
elements
\beq
\weyl_\star(\mk)=\e^{\ii k^a\,\X_a}
\label{Weylweylop}\eeq
with $\mk\in\mbbV$ and $\X_a\in\mfs$.

The Weyl system $(\mbbV,\compc,\weyl_\star,\omega_\star)$ can be used
to generate the other Weyl systems that we have
found~\cite{ALZ1}. From (\ref{1ststudyexpl}) and
(\ref{Weylgpprodexpl}) one has the identity
\beq
\weyl_*(j,\mbp^\pm)=\Omega_\star\left(
\e^{\ii(\mbf p^+\cdot\overline{\mz}
+\mbp^-\cdot\mz)}\star\e^{\ii j\, x^-}\right)
\label{weylTOWeylDelta}\eeq
which implies
that the time-ordered star-product $*$ can be expressed
by means of a choice of different Weyl system generating the product
$\star$. Since $\Omega_\star$ is an algebra isomorphism, one has
\beq
\weyl_*(j,\mbp^\pm)=\weyl_\star(0,\mbp^\pm)
\cdot\weyl_\star(j,\mbf0) \ .
\label{weylTOWeylprods}\eeq
This explicit relationship between the Weyl systems for the
star-products $*$ and $\star$ is another formulation of the statement
of their cohomological equivalence, as established by other means in
Section~\ref{WOP}. Similarly, the symmetric time-ordered star-product
$\bullet$ can be expressed in terms of $\star$ through the identity
\beq
\weyl_\bullet(j,\mbp^\pm)=\Omega_\star\left(
\e^{\frac\ii2\,j\,x^-}\star\e^{\ii(\mbp^+\cdot\overline{\mz}
+\mbp^-\cdot\mz)}\star\e^{\frac\ii2\,j\,x^-}\right) \ ,
\label{WeylTOsymWeylDelta}\eeq
which implies the relationship
\beq
\weyl_\bullet\bigl(j,\mbp^\pm\bigr)=\weyl_\star
\bigl(\mbox{$\frac j2$}
,\mbf0\bigr)\cdot\weyl_\star\bigl(0,\mbp^\pm\bigr)
\cdot\weyl_\star\bigl(\mbox{$\frac j2$},\mbf0\bigr)
\label{WeylTOsymWeylprods}\eeq
between the corresponding Weyl systems. This shows explicitly that the
star-products $\bullet$ and $\star$ are also equivalent.

\newsection{Twisted Isometries\label{Coprod}}

We will now start working our way towards the explicit construction of
the geometric quantities required to define field
theories on the noncommutative plane wave $\NW_6$. We will begin with
a systematic construction of derivative operators on the present
noncommutative geometry, which will be used later on to write down
kinetic terms for scalar field actions. In this section we will study
some of the basic spacetime symmetries of the star-products that we
constructed in Section~\ref{StarProds}, as they are directly related
to the actions of derivations on the noncommutative algebras of
functions.

Classically, the isometry group of the gravitational wave $\NW_6$ is
the group $\mcN_{\rm L}\times\mcN_{\rm R}$ induced by the left and right
regular actions of the Lie group $\mcN$ on itself. The corresponding
Killing vectors live in the 11-dimensional Lie algebra $\mfg:=\mfn_{\rm
  L}\oplus\mfn_{\rm R}$ (The left and right actions generated by the
central element $\T$ coincide). This isometry group contains an ${\rm
  SO}(4)$ subgroup acting by rotations in the transverse space
$\mz\in\complex^2\cong\real^4$, which is broken to ${\rm U}(2)$ by the
Neveu-Schwarz background (\ref{NS2formBrink}). This symmetry can be
restored upon quantization by instead letting the generators of $\mfg$
act in a twisted fashion~\cite{CPT1,CKNT1,Wess1}, as we now proceed to
describe.

The action of an element $\nabla\in U(\mfg)$ as an algebra
automorphism $\CC^\infty(\mfn^{\vee\,})\to\CC^\infty(\mfn^{\vee\,})$ will be
denoted
$f\mapsto\nabla\triangleright f$. The universal enveloping algebra
$U(\mfg)$ is given the structure of a cocommutative bialgebra by
introducing the ``trivial'' coproduct $\Delta:U(\mfg)\to
U(\mfg)\otimes U(\mfg)$ defined by the homomorphism
\beq
\Delta(\nabla)=\nabla\otimes1+1\otimes\nabla \ ,
\label{trivialcoprod}\eeq
which generates the action of $U(\mfg)$ on the tensor product
$\CC^\infty(\mfn^{\vee\,})\otimes\CC^\infty(\mfn^{\vee\,})$. Since $\nabla$ is
an
automorphism of $\CC^\infty(\mfn^{\vee\,})$, the  action of the coproduct is
compatible with the pointwise (commutative) product of functions
$\mu:\CC^\infty(\mfn^{\vee\,})\otimes\CC^\infty(\mfn^{\vee\,})\to
\CC^\infty(\mfn^{\vee\,})$ in the sense that
\beq
\nabla\triangleright\mu(f\otimes g)=\mu\circ\Delta(\nabla)
\triangleright(f\otimes g) \ .
\label{commcoprodcomp}\eeq
For example, the standard action of spacetime translations is given by
\beq
\partial^a\triangleright f=\partial^af
\label{commtranslaction}\eeq
for which (\ref{commcoprodcomp}) becomes the classical symmetric
Leibniz rule.

Let us now pass to a noncommutative deformation of the algebra of
functions on $\NW_6$ via a quantization map
$\Omega:\CC^\infty(\mfn^{\vee\,})\to\overline{U(\mfn)^\complex}$
corresponding to a specific star-product $\star$ on
$\CC^\infty(\mfn^{\vee\,})$ (or equivalently a specific operator ordering in
$U(\mfn)$). This isomorphism can be used to induce an action of
$U(\mfg)$ on the algebra $\overline{U(\mfn)^\complex}$ through
\beq
\Omega(\nabla_\star)\triangleright\Omega(f):=
\Omega(\nabla\triangleright f) \ ,
\label{nablastar}\eeq
which defines a set of quantized operators
$\nabla_\star=\nabla+O(\theta):\CC^\infty(\mfn^{\vee\,})\to\CC^\infty(\mfn^{\vee\,})$.
However, the bialgebra $U(\mfg)$ will no longer generate automorphisms
with respect to the noncommutative star-product on
$\CC^\infty(\mfn^{\vee\,})$. It will only do so if its coproduct can be
deformed to a non-cocommutative one $\Delta_\star=\Delta+O(\theta)$
such that the covariance condition
\beq
\nabla_\star\triangleright\mu_\star(f\otimes g)=\mu_\star\circ
\Delta_\star(\nabla_\star)\triangleright(f\otimes g)
\label{NCcoprodcomp}\eeq
is satisfied, where $\mu_\star(f\otimes g):=f\star g$. This deformation is
constructed by writing the star-product $f\star
g=\hat{\mathcal D}(f,g)$ in terms of a bi-differential operator as in
(\ref{fstargbidiff}) or (\ref{bidiffweyl}) to define an invertible
abelian Drinfeld twist element~\cite{Resh1} $\hat{\mathcal F}_\star\in
\overline{U(\mfg)^\complex}\otimes\overline{U(\mfg)^\complex}$ through
\beq
f\star g=\mu\circ\hat{\mathcal F}{}_\star^{-1}\triangleright(f\otimes
g) \ .
\label{Dtwistdef}\eeq
It obeys the cocycle condition
\beq
(\hat{\mathcal F}_\star\otimes1)\,(\Delta\otimes1)\,\hat
{\mathcal F}_\star=(1\otimes\hat{\mathcal F}_\star)\,
(\Delta\otimes1)\,\hat{\mathcal F}_\star
\label{twistcocycle}\eeq
and defines the twisted coproduct through
\beq
\Delta_\star:=\hat{\mathcal F}_\star^{~}\circ\Delta\circ
\hat{\mathcal F}{}_\star^{-1} \ ,
\label{Deltastardef}\eeq
where $(f\otimes g)\circ(f'\otimes g'\,):=f\,f'\otimes g\,g'$. This new
coproduct obeys the requisite coassociativity condition
$(\Delta_\star\otimes\one)\circ\Delta_\star=(\one\otimes\Delta_\star)
\circ\Delta_\star$. The important property of the twist element
$\hat{\mathcal F}_\star$ is that it modifies only the coproduct on the
bialgebra $U(\mfg)$, while leaving the original product structure
(inherited from the Lie algebra $\mfg=\mfn_{\rm L}\oplus\mfn_{\rm
  R}$) unchanged.

As an example, let us illustrate how to compute the twisting of the
quantized translation generators by the noncommutative geometry of $\NW_6$. For
this, we introduce a Weyl system $(\mbbV,\comp,\weyl,\omega)$
corresponding to the chosen star-product $\star$. With the same
notations as in the previous section, for $a=1,\dots,5$ we may use
(\ref{Weylcomprule}), (\ref{involdef}) with $\Pi=\Omega$, and
(\ref{nablastar}) with $\nabla=\partial^a$ to compute
\bea
\Omega\bigl(\partial_\star^a\bigr)\triangleright\Omega\bigl(
\e^{\ii\mk\cdot\mx}\bigr)\cdot\Omega\bigl(\e^{\ii\mk'\cdot\mx}
\bigr)&=&\Omega\bigl(\partial_\star^a\bigr)\triangleright
\e^{\frac\ii2\,\omega(\mk,\mk'\,)\,\T}~\cdot~\Omega\bigl(
\e^{\ii(\mk\comp\mk'\,)\cdot\mx}\bigr) \nonumber\\[4pt] &=&
\ii~\e^{\frac\ii2\,\omega(\mk,\mk'\,)\,\T}~\cdot~\Omega
\bigl((\mk\comp\mk'\,)^a~\e^{\ii(\mk\comp\mk'\,)\cdot\mx}\bigr)
\nonumber\\[4pt] &=& \ii~\mbox{$\sum\limits_i$}~\Omega\bigl(d^a_{(1)\,i}(
-\ii\mdell_\star)\bigr)\triangleright\Omega\bigl(\e^{\ii\mk\cdot\mx}
\bigr)\nonumber\\ && \qquad\qquad
\cdot~\Omega\bigl(d^a_{(2)\,i}(-\ii\mdell_\star)\bigr)
\triangleright\Omega\bigl(\e^{\ii\mk'\cdot\mx}\bigr) \ ,
\label{partialstarderiv}\eea
where we have assumed that the group composition law of the Weyl
system has an expansion of the form
$(\mk\comp\mk'\,)^a:=\sum_i\,d^a_{(1)\,i}(\mk)\,d^a_{(2)\,i}(\mk'\,)$. From
the covariance condition (\ref{NCcoprodcomp}) it then follows that
the twisted coproduct assumes a Sweedler form
\beq
\Delta_\star\left(\partial_\star^a\right)=\ii~
\mbox{$\sum\limits_i$}~d^a_{(1)\,i}(-\ii\mdell_\star)\otimes
d^a_{(2)\,i}(-\ii\mdell_\star) \ .
\label{Deltastarexpl}\eeq
Analogously, if we assume that the group cocycle of the Weyl system
admits an expansion of the form
$\omega(\mk,\mk'\,):=\sum_i\,w^i_{(1)}(\mk)\,w^i_{(2)}(\mk'\,)$, then
a similar calculation gives the twisted coproduct of the quantized
plane wave time derivative as
\beq
\Delta_\star\left(\partial_+^\star\right)=\partial_+^\star
\otimes1+1\otimes\partial_+^\star-\mbox{$\frac12\,\sum\limits_i$}~
w^i_{(1)}(-\ii\mdell_\star)\otimes w^i_{(2)}(-\ii\mdell_\star) \ .
\label{Deltastartime}\eeq
Note that now the corresponding Leibniz rules (\ref{NCcoprodcomp}) are
no longer the usual ones associated with the product $\star$ but are
the deformed, generically non-symmetric ones given by
\bea
\partial^a_\star\triangleright(f\star g)&=&\ii~\mbox{$\sum\limits_i$}\,
\bigl(d^a_{(1)\,i}(-\ii\mdell_\star)\triangleright f\bigr)~\star~\bigl(
d^a_{(2)\,i}(-\ii\mdell_\star)\triangleright g\bigr) \ , \nonumber\\[4pt]
\partial_+^\star\triangleright(f\star g)&=&\left(\partial_+^\star
\triangleright f\right)\star g+f\star\left(\partial_+^\star
\triangleright g\right)-\mbox{$\frac12\,\sum\limits_i$}\,
\bigl(w^i_{(1)}(-\ii\mdell_\star)\triangleright f\bigr)~\star~
\bigl(w^i_{(2)}(-\ii\mdell_\star)\triangleright g\bigr) \nonumber\\ &&
\label{defLeibniz}\eea
arising from the twisting of the coproduct. Thus these derivatives do
{\it not} define derivations of the noncommutative algebra of
functions, but rather implement the twisting of isometries of flat
space appropriate to the plane wave
geometry~\cite{PK1,CFS1,BlauOL1,HSz1}.

In the language of quantum groups~\cite{QG1}, the twisted isometry
group of the spacetime $\NW_6$ coincides with the quantum double of the
cocommutative Hopf algebra $U(\mfn)$. The antipode
${S}_\star:U(\mfg)\to U(\mfg)$ of the given non-cocommutative Hopf
algebra structure on the bialgebra $U(\mfg)$ gives the dual action of
the isometries of the noncommutative plane wave and provides the
analog of inversion of isometry group elements. This analogy is made
precise by computing ${S}_\star$ from the group inverses
$\underline{\mk}$ of elements $\mk\in\mbbV$ of the corresponding Weyl
system. Symbolically, one has
${S}_\star(\mdell_\star)=\underline{\mdell_\star}$. In particular, if
$\underline{\mk}=-\mk$ (as in the case of our symmetric star-products)
then ${S}_\star(\partial_\star^a)=-\partial_\star^a$ and the action
of the antipode is trivial. In all three instances the counit
$\varepsilon_\star:U(\mfg)\to\complex$ describes the action on the
trivial representation as $\varepsilon_\star(\partial_\star^a)=0$, and
it obeys the compatibility condition
\beq
(\varepsilon_\star\otimes1)\,\hat{\cal F}_\star~=~1~=~
(1\otimes\varepsilon_\star)\,\hat{\cal F}_\star
\label{counitcond}\eeq
with the Drinfeld twist. In what follows we will only require the
underlying bialgebra structure of $U(\mfg)$. The compatibility
condition (\ref{NCcoprodcomp}) means that the action of $U(\mfg)$ on
$\CC^\infty(\mfn^{\vee\,})$ defines quantum isometries of the noncommutative
pp-wave, in that the star-product is an intertwiner and the
noncommutative algebra of functions is covariant with respect to the
action of the quantum group.

The generic non-triviality of the twisted coproducts
(\ref{Deltastarexpl}) and (\ref{Deltastartime}) is consistent with and
extends the fact that generic translations are not classically
isometries of the plane wave geometry, but rather only appropriate
twisted versions are~\cite{PK1,CFS1,BlauOL1,HSz1}. Similar
computations can also be carried through
for the remaining five isometry generators of $\mfg$ and correspond to
the right-acting counterparts of the derivatives above, giving the full
action of the noncommutative isometry group on $\NW_6$. We shall not
display these formulas here. In the next section we will explicitly
construct the quantized derivative operators $\partial_\star^a$ and
$\partial_+^\star$ above. We now proceed to list the coproducts
corresponding to our three star-products.

\subsection{Time Ordering \label{TOcoprod}}

The Drinfeld twist $\hat{\mathcal F}_*$ for the time-ordered
star-product is the inverse of the exponential operator appearing in
(\ref{TOstargen}). Following the general prescription given above,
from the group composition law (\ref{TOcomplaw}) of the corresponding
Weyl system we deduce the time-ordered coproducts
\bea
\Delta_*\left(\partial_-^*\right)&=&\partial_-^*\otimes1+
1\otimes\partial_-^* \ , \nonumber\\
\Delta_*\left(\partial^i_*\right)&=&\partial^i_*\otimes1+
\e^{\ii\theta\,\partial_-^*}\otimes\partial^i_* \ , \nonumber\\
\Delta_*\left(\,\overline{\partial}{}^{\,i}_*\right)&=&
\overline{\partial}{}^{\,i}_*\otimes1+\e^{-\ii\theta\,\partial_-^*}
\otimes\overline{\partial}{}^{\,i}_* \ ,
\label{TOcoprods}\eea
while from the group cocycle (\ref{TOcocycle}) we obtain
\beq
\Delta_*\left(\partial_+^*\right)=\partial_+^*\otimes1+
1\otimes\partial_+^*+\theta~\e^{-\ii\theta\,
\partial_-^*}\,\mdell_*{}^\top\otimes\overline{\mdell}_*-\theta~
\e^{\ii\theta\,\partial_-^*}\,
\overline{\mdell}_*{}^\top\otimes\mdell_* \ .
\label{TOcoprodtime}\eeq
The corresponding Leibniz rules read
\bea
\partial_-^*\triangleright(f*g)&=&\left(\partial_-^*\triangleright
f\right)*g+f*\left(\partial_-^*\triangleright g\right) \ , \nonumber\\
\partial_+^*\triangleright(f*g)&=&\left(\partial_+^*\triangleright
f\right)*g+f*\left(\partial_+^*\triangleright g\right)\nonumber\\ &&
+\,\theta\,\bigl(\e^{-\ii\theta\,\partial_-^*}\,\mdell_*{}^\top
\triangleright f\bigr)~*~\left(\,\overline{\mdell}_*
\triangleright g\right)-\theta\,\bigl(\e^{\ii\theta\,\partial_-^*}\,
\overline{\mdell}_*{}^\top
\triangleright f\bigr)~*~\left(\mdell_*\triangleright g\right) \ ,
\nonumber\\ \partial^i_*\triangleright(f*g)&=&\left(\partial^i_*
\triangleright f\right)*g+\bigl(\e^{\ii\theta\,\partial_-^*}
\triangleright f\bigr)*\left(\partial^i_*\triangleright g\right) \ ,
\nonumber\\ \overline{\partial}{}^{\,i}_*
\triangleright(f*g)&=&\left(\,\overline{\partial}{}^{\,i}_*
\triangleright f\right)*g+\bigl(\e^{-\ii\theta\,\partial_-^*}
\triangleright f\bigr)*\left(\,\overline{\partial}{}^{\,i}_*
\triangleright g\right) \ .
\label{TOLeibniz}\eea

\subsection{Symmetric Time Ordering \label{STOcoprod}}

The Drinfeld twist $\hat{\mathcal F}_\bullet$ associated to the
symmetric time-ordered star-product is given by the inverse of the
exponential operator in (\ref{TOsymstargen}). From the group
composition law (\ref{TOsymcomplaw}) of the corresponding Weyl system
we deduce the symmetric time-ordered coproducts
\bea
\Delta_\bullet\left(\partial_-^\bullet\right)&=&\partial_-^\bullet
\otimes1+1\otimes\partial_-^\bullet \ , \nonumber\\
\Delta_\bullet\left(\partial^i_\bullet\right)&=&\partial^i_\bullet\otimes
\e^{-\frac{\ii\theta}2\,\partial_-^\bullet}+\e^{\frac{\ii\theta}2\,
\partial_-^\bullet}\otimes\partial^i_\bullet \ , \nonumber\\
\Delta_\bullet\left(\,\overline{\partial}{}^{\,i}_\bullet\right)&=&
\overline{\partial}{}^{\,i}_\bullet\otimes
\e^{\frac{\ii\theta}2\,\partial_-^\bullet}+\e^{-\frac{\ii\theta}2\,
\partial_-^\bullet}\otimes\overline{\partial}{}^{\,i}_\bullet \ ,
\label{STOcoprods}\eeq
while from the group cocycle (\ref{TOsymcocycle}) we find
\bea
\Delta_\bullet\left(\partial_+^\bullet\right)&=&\partial_+^\bullet\otimes1+
1\otimes\partial_+^\bullet\nonumber\\ &&
+\,\theta~\e^{-\frac{\ii\theta}2\,
\partial_-^\bullet}\,\mdell_\bullet{}^\top\otimes
\e^{-\frac{\ii\theta}2\,\partial_-^\bullet}\,\overline{\mdell}_\bullet-
\theta~\e^{\frac{\ii\theta}2\,\partial_-^\bullet}\,
\overline{\mdell}_\bullet{}^\top\otimes\e^{\frac{\ii\theta}2\,
\partial_-^\bullet}\,\mdell_\bullet \ .
\label{STOcoprodtime}\eea
The corresponding Leibniz rules are given by
\bea
\partial_-^\bullet\triangleright(f\bullet g)&=&
\left(\partial_-^\bullet\triangleright f\right)\bullet g+
f\bullet\left(\partial_-^\bullet\triangleright g\right) \ ,
\nonumber\\ \partial_+^\bullet\triangleright(f\bullet g)&=&
\left(\partial_+^\bullet\triangleright f\right)\bullet g+
f\bullet\left(\partial_+^\bullet\triangleright g\right)
+\theta\,\bigl(\e^{-\frac{\ii\theta}2\,\partial_-^\bullet}\,
\mdell_\bullet{}^\top\triangleright f\bigr)~\bullet~\bigl(
\e^{-\frac{\ii\theta}2\,\partial_-^\bullet}\,
\overline{\mdell}_\bullet\triangleright g\bigr) \nonumber\\ &&
\qquad\qquad\qquad\qquad\qquad\qquad -\,
\theta\,\bigl(\e^{\frac{\ii\theta}2\,\partial_-^\bullet}\,
\overline{\mdell}_\bullet{}^\top\triangleright f\bigr)~\bullet~
\bigl(\e^{\frac{\ii\theta}2\,\partial_-^\bullet}\,\mdell_\bullet
\triangleright g\bigr) \ , \nonumber\\
\partial^i_\bullet\triangleright(f\bullet g)&=&\left(\partial^i_\bullet
\triangleright f\right)\bullet\bigl(\e^{-\frac{\ii\theta}2\,
\partial_-^\bullet}\triangleright g\bigr)+\bigl(
\e^{\frac{\ii\theta}2\,\partial_-^\bullet}\triangleright f
\bigr)\bullet\left(\partial^i_\bullet\triangleright g\right) \ ,
\nonumber\\ \overline{\partial}{}^{\,i}_\bullet\triangleright(f\bullet g)
&=&\left(\,\overline{\partial}{}^{\,i}_\bullet
\triangleright f\right)\bullet\bigl(\e^{\frac{\ii\theta}2\,
\partial_-^\bullet}\triangleright g\bigr)+\bigl(
\e^{-\frac{\ii\theta}2\,\partial_-^\bullet}\triangleright f
\bigr)\bullet\left(\,\overline{\partial}{}^{\,i}_\bullet\triangleright g\right)
\ .
\label{STOLeibniz}\eea

\subsection{Weyl Ordering \label{WOcoprod}}

Finally, for the Weyl-ordered star-product (\ref{Weylstargen}) we read
off the twist element $\hat{\mathcal F}_\star$ in the standard way,
and use the associated group composition law (\ref{Weylcomplaw}) to
write down the coproducts
\bea
\Delta_\star\left(\partial_-^\star\right)&=&\partial_-^\star\otimes1+
1\otimes\partial_-^\star \ , \nonumber\\
\Delta_\star\left(\partial^i_\star\right)&=&\mbox{$\frac{\phi_{-\theta}\left(\ii
\partial_-^\star\right)\,\partial^i_\star\otimes1+\e^{\ii\theta\,\partial_-^\star}
\otimes\phi_{-\theta}\left(\ii\partial_-^\star\right)\,\partial^i_\star}
{\phi_{-\theta}\left(\ii\partial_-^\star\otimes1+1\otimes\ii
\partial_-^\star\right)}$} \ , \nonumber\\
\Delta_\star\left(\,\overline{\partial}{}^{\,i}_\star\right)&=&
\mbox{$\frac{\phi_{\theta}\left(\ii
\partial_-^\star\right)\,\overline{\partial}{}^{\,i}_\star\otimes1+
\e^{-\ii\theta\,\partial_-^\star}
\otimes\phi_{\theta}\left(\ii\partial_-^\star\right)\,
\overline{\partial}{}^{\,i}_\star}
{\phi_{\theta}\left(\ii\partial_-^\star\otimes1+1\otimes\ii
\partial_-^\star\right)}$} \ .
\label{Weylcoprods}\eea
The remaining coproduct may be determined from the cocycle
(\ref{Weylcocycle}) as
\bea
\Delta_\star\left(\partial_+^\star\right)&=&\partial_+^\star\otimes1+
1\otimes\partial_+^\star \nonumber\\ &&+\,2\ii\theta\,\Bigl[
\phi_\theta\left(\ii\partial_-^\star\right)\,\mdell_\star
{}^\top\otimes\phi_\theta\left(\ii\partial_-^\star\right)\,
\overline{\mdell}_\star-\phi_{-\theta}\left(\ii\partial_-^\star
\right)\,\overline{\mdell}_\star
{}^\top\otimes\phi_{-\theta}\left(\ii\partial_-^\star\right)\,
\mdell_\star\Bigr.
\nonumber\\ && \qquad\quad
+\,\bigl(\gamma_\theta(\ii\partial_-^\star)\otimes1-
\gamma_\theta(\ii\partial_-^\star\otimes1+
1\otimes\ii\partial_-^\star)\bigr)\bigl(\phi_\theta(\ii\partial_-^\star)
\,\phi_{-\theta}(\ii\partial_-^\star)\,\overline{\mdell}_\star
\cdot\mdell_\star\otimes1\bigr) \nonumber\\ &&\qquad\quad
+\,\bigl(1\otimes\gamma_\theta(\ii\partial_-^\star)-
\gamma_\theta(\ii\partial_-^\star\otimes1+
1\otimes\ii\partial_-^\star)\bigr)\bigl(1\otimes
\phi_\theta(\ii\partial_-^\star)
\,\phi_{-\theta}(\ii\partial_-^\star)\,\overline{\mdell}_\star
\cdot\mdell_\star\bigr) \nonumber\\ &&\qquad\quad -\Bigl.\gamma_\theta\left(
\ii\partial_-^\star\otimes1+1\otimes\ii\partial_-^\star\right)
\bigl(\e^{-\ii\theta\,\partial_-^\star}\,\phi_{-\theta}(\ii
\partial_-^\star)\,\mdell_\star{}^\top\otimes\phi_\theta(\ii
\partial_-^\star)\,\overline{\mdell}_\star\bigr.\nonumber\\ &&
\qquad\qquad\qquad\qquad\qquad\qquad\qquad\quad+\bigl.\e^{\ii\theta\,
\partial_-^\star}\,\phi_{\theta}(\ii
\partial_-^\star)\,\overline{\mdell}_\star{}^\top\otimes
\phi_{-\theta}(\ii\partial_-^\star)\,\mdell_\star\bigr)\Bigr] \ .
\nonumber\\ &&
\label{Weylcoprodtime}\eea
In (\ref{Weylcoprods}) and (\ref{Weylcoprodtime}) the functionals of
the derivative operator
$\ii\partial_-^\star\otimes1+1\otimes\ii\partial_-^\star$ are
understood as usual in terms of the power series expansions given in
Section~\ref{WOP}. This leads to the corresponding Leibniz rules
\bea
\partial_-^\star\triangleright(f\star g)&=&\left(\partial_-^\star
\triangleright f\right)\star g+f\star\left(\partial_-^\star
\triangleright g\right) \ , \nonumber\\
\partial_+^\star\triangleright (f\star g)&=&\left(\partial_+^\star
\triangleright f\right)\star g+f\star\left(\partial_+^\star
\triangleright g\right) \nonumber\\ &&+\,2\ii\theta\left\{
\Bigl(\mbox{$\frac{(1-\e^{-\ii\theta\,\partial_-^\star})\,
\mdell_\star{}^\top}{\ii\theta\,\partial_-^\star}$}\triangleright f
\Bigr)~\star~\Bigl(\mbox{$\frac{(1-\e^{-\ii\theta\,\partial_-^\star})\,
\overline{\mdell}_\star}{\ii\theta\,\partial_-^\star}$}\triangleright
g\Bigr)\right. \nonumber\\ && \qquad\qquad
-\,\Bigl(\mbox{$\frac{(1-\e^{\ii\theta\,
\partial_-^\star})\,\overline{\mdell}_\star{}^\top}
{\ii\theta\,\partial_-^\star}$}\triangleright f
\Bigr)~\star~\Bigl(\mbox{$\frac{(1-\e^{\ii\theta\,\partial_-^\star})\,
\mdell_\star}{\ii\theta\,\partial_-^\star}$}\triangleright
g\Bigr)\nonumber\\ &&\qquad\qquad
+\,\Bigl(\mbox{$\Bigl[\frac12+\frac{(1+\ii\theta\,
\partial_-^\star)~\e^{-\ii\theta\,\partial_-^\star}-1}{(\e^{-\ii\theta\,
\partial_-^\star}-1)^2}\Bigr]\,\frac{\sin^2(\frac\theta2\,\partial_-^\star)\,
\overline{\mdell}_\star\cdot\mdell_\star}{(\theta\,\partial_-^\star)^2}$}
\triangleright f\Bigr)~\star~g\nonumber\\ &&\qquad\qquad+\,f~\star~
\Bigl(\mbox{$\Bigl[\frac12+\frac{(1+\ii\theta\,
\partial_-^\star)~\e^{-\ii\theta\,\partial_-^\star}-1}{(\e^{-\ii\theta\,
\partial_-^\star}-1)^2}\Bigr]\,\frac{\sin^2(\frac\theta2\,\partial_-^\star)\,
\overline{\mdell}_\star\cdot\mdell_\star}{(\theta\,\partial_-^\star)^2}$}
\triangleright g\Bigr) \nonumber\\ && +\,\sum_{n=1}^\infty~
\sum_{k=0}^n\,\frac{B_{n+1}\,(-\ii\theta)^{n-2}}{k!\,(n-k)!}\,
\left[\bigl((\partial_-^\star)^{n-k-2}\,\sin^2(\mbox{$\frac\theta2$}\,
\partial_-^\star)\,\overline{\mdell}_\star\cdot\mdell_\star
\triangleright f\bigr)~\star~\bigl((\partial_-^\star)^k
\triangleright g\bigr)\right.\nonumber\\ && \qquad +\,
\bigl((\partial_-^\star)^{n-k}\triangleright
f\bigr)~\star~\bigl((\partial_-^\star)^{k-2}\,\sin^2(\mbox{$\frac\theta2$}\,
\partial_-^\star)\,\overline{\mdell}_\star\cdot\mdell_\star
\triangleright g\bigr) \nonumber\\ && \qquad -\,
\bigl((\e^{-\ii\theta\,\partial_-^\star}-1)\,(\partial_-^\star)^{n-k-1}
\,\mdell_\star{}^\top\triangleright f\bigr)~\star~\bigl(
(\e^{-\ii\theta\,\partial_-^\star}-1)\,(\partial_-^\star)^{k-1}
\,\overline{\mdell}_\star\triangleright g\bigr) \nonumber\\ &&
\qquad -\left.
\bigl((\e^{\ii\theta\,\partial_-^\star}-1)\,(\partial_-^\star)^{n-k-1}
\,\overline{\mdell}_\star{}^\top\triangleright f\bigr)~\star~\bigl(
(\e^{\ii\theta\,\partial_-^\star}-1)\,(\partial_-^\star)^{k-1}
\,\mdell_\star\triangleright g\bigr)\right] \ , \nonumber\\
\partial^i_\star\triangleright(f\star g)&=&\sum_{n=0}^\infty~
\sum_{k=0}^n\,\frac{B_n\,(\ii\theta)^{n-1}}{k!\,(n-k)!}\,\left[
\bigl((\e^{\ii\theta\,\partial_-^\star}-1)\,
(\partial_-^\star)^{n-k-1}\,\partial^i_\star\triangleright f\bigr)~\star~
\bigl((\partial_-^\star)^k\triangleright g\bigr)\right.\nonumber\\ &&
\qquad\qquad\qquad\quad
+\left.\bigl(\e^{\ii\theta\,\partial_-^\star}\,(\partial_-^\star)^{n-k}
\triangleright f\bigr)~\star~\bigl((\e^{\ii\theta\,\partial_-^\star}-1)
\,(\partial_-^\star)^{k-1}\,\partial^i_\star\triangleright g\bigr)\right]
\ , \nonumber\\\overline{\partial}{}^{\,i}_\star\triangleright(f\star g)&=&
\sum_{n=0}^\infty~\sum_{k=0}^n\,\frac{B_n\,(-\ii\theta)^{n-1}}{k!\,(n-k)!}
\,\left[\bigl((\e^{-\ii\theta\,\partial_-^\star}-1)\,
(\partial_-^\star)^{n-k-1}\,\overline{\partial}{}^{\,i}_\star
\triangleright f\bigr)~\star~
\bigl((\partial_-^\star)^k\triangleright g\bigr)\right.\nonumber\\ &&
\qquad\qquad\qquad\quad
+\left.\bigl(\e^{-\ii\theta\,\partial_-^\star}\,(\partial_-^\star)^{n-k}
\triangleright f\bigr)~\star~\bigl((\e^{-\ii\theta\,\partial_-^\star}-1)
\,(\partial_-^\star)^{k-1}\,\overline{\partial}{}^{\,i}_\star
\triangleright g\bigr)\right] \ .
 \nonumber\\ &&
\label{WeylLeibniz}\eea

Note that a common feature to all three deformations is that the coproduct
of the quantization of the light-cone position translation generator
$\partial_-$ coincides with the trivial one (\ref{trivialcoprod}), and
thereby yields the standard symmetric Leibniz rule with respect to the
pertinent star-product. This owes to the fact that the action of
$\partial_-$ on the spacetime $\NW_6$ corresponds to the commutative
action of the central Lie algebra generator $\T$, whose left and right actions
coincide. In the next section we shall see that the action of the
quantized translations in $x^-$ on $\CC^\infty(\mfn^{\vee\,})$ coincides with
the standard commutative action (\ref{commtranslaction}). This is
consistent with the fact that all frames of reference for the
spacetime $\NW_6$ possess an $x^-$-translational symmetry, while
translational symmetries in the other coordinates depend crucially on
the frame and generally need to be twisted in order to generate an
isometry of $\NW_6$. Notice also that ordinary time translation
invariance is always broken by the time-dependent Neveu-Schwarz
background (\ref{NS2formBrink}).

\newsection{Derivative Operators \label{Derivatives}}

In this section we will systematically construct a set of quantized
derivative operators $\partial^a_\star$, $a=1,\dots,6$ satisfying the
conditions of the previous section. In general, there is no unique way
to build up such derivatives. To this end, we will impose some weak
conditions, namely that the quantized derivatives be deformations of
ordinary derivatives, $\partial_\star^a=\partial^a+O(\theta)$, and
that they commute among themselves,
$[\partial_\star^a,\partial_\star^b]_\star=0$. The latter condition is
understood as a requirement for the iterated action of the derivatives
on functions $f\in\CC^\infty(\mfn^{\vee\,})$,
$[\partial_\star^a,\partial_\star^b]_\star\triangleright f=0$ or
equivalently
\beq
\partial_\star^a\triangleright\bigl(\partial_\star^b\triangleright f
\bigr)=\partial_\star^b\triangleright\bigl(\partial_\star^a\triangleright
f\bigr) \ .
\label{derivcommute}\eeq
For the former condition, the simplest consistent choice is to assume
a linear derivative deformation on the coordinate functions,
$[\partial_\star^a,x_b]_\star=
\delta^a_{~b}+\ii\theta\,\rho^a_{~bc}\,\partial^c_\star$, which is
understood as the requirement
\beq
\left[\partial_\star^a\,,\,x_b\right]_\star\triangleright f:=
\partial^a_\star\triangleright\left(x_b\star f\right)-
x_b\star\left(\partial_\star^a\triangleright f\right)=
\delta^a_{~b}\,f+\ii\theta\,\rho^a_{~bc}\,\partial^c_\star
\triangleright f \ .
\label{dxreq}\eeq
A set of necessary conditions on the constant tensors
$\rho^a_{~bc}\in\real$ may be derived by demanding consistency of the
derivatives with the original star-commutators of coordinates
(\ref{xaxbstarcomm}). Applying the Jacobi identity for the
star-commutators between $\partial^a_\star$, $x_b$ and $x_c$ leads to
the relations
\bea
\rho^a_{~bc}-\rho^a_{~cb}&=&C_{bc}^{~~a} \ , \nonumber\\
\rho^a_{~bc}\,\rho^c_{~de}-\rho^a_{~dc}\,\rho^c_{~be}&=&
C_{bd}^{~~c}\,\rho^a_{~ce} \ .
\label{rhoCrels}\eeq

With these requirements we now seek to find quantized derivative
operators $\partial_\star^a$ as functionals of ordinary derivatives
$\partial^a$ acting on $\CC^\infty(\mfn^{\vee\,})$ as in
(\ref{commtranslaction}). However, there are (uncountably) infinitely
many solutions $\rho^a_{~bc}$ obeying (\ref{rhoCrels})~\cite{DMT1} with
$C_{ab}^{~~c}$ the structure constants of the Lie algebra $\mfn$ given
by (\ref{NW4algdef}). We will choose the simplest consistent one
defined by the star-commutators
\begin{align}
  \nonumber
  \left[\partial_-^\star\,,\,x^-\right]_\star&=1 \ ,
  &\left[\partial_+^\star\,,\,x^-\right]_\star&=0 \ ,
  &\left[\partial_\star^i\,,\,x^-\right]_\star&=-\ii\theta\,\partial_\star^i \
,
  &\left[\,\overline{\partial}{}_\star^{\,i}\,,\,x^-\right]_\star&=
  \ii\theta\,\overline{\partial}{}_\star^{\,i} \ ,
  \\ \nonumber
  \left[\partial_-^\star\,,\,x^+\right]_\star&=0 \ ,
  &\left[\partial_+^\star\,,\,x^+\right]_\star &=1 \ ,
  &\left[\partial_\star^i\,,\,x^+\right]_\star&=0 \ ,
  &\left[\,\overline{\partial}{}_\star^{\,i}\,,\,x^+\right]_\star&=0 \ ,
  \\ \nonumber
  \left[\partial_-^\star\,,\,z_i\right]_\star&=0 \ ,
  &\left[\partial_+^\star\,,\,z_i\right]_\star&=-\ii\theta\,
  \overline{\partial}{}^{\,i}_\star \ ,
  &\left[\partial_\star^i\,,\,z_j\right]_\star&=\delta^i_{~j} \ ,
  &\left[\,\overline{\partial}{}_\star^{\,i}\,,\,z_j\right]_\star&=0 \ ,
  \\
  \left[\partial_-^\star\,,\,\overline{z}_i\right]_\star&=0  \ ,
  &\left[\partial_+^\star\,,\,\overline{z}_i\right]_\star&=
  \ii\theta\,\partial_\star^i \ ,
  &\left[\partial_\star^i\,,\,\overline{z}_j\right]_\star&=0 \ ,
  &\left[\,\overline{\partial}{}_\star^{\,i}\,,\,\overline{z}_j\right]_\star&
  =\delta^i_{~j} \ ,
\label{eq:rho:nw4}\end{align}
whose $O(\theta)$ parts mimick the structure of the Lie brackets
(\ref{NW4algdef}). This choice ensures that the derivatives
$\partial_\star^a$ will generate the isometries appropriate to the
quantization of the curved spacetime $\NW_6$. All other admissible
choices for $\rho^a_{~bc}$ can be mapped into those given by
(\ref{eq:rho:nw4}) via non-linear redefinitions of the derivative
operators $\partial^a_\star$~\cite{DMT1}. It is
important to realize that the quantized derivatives do not generally
obey the classical Leibniz rule,
i.e. $\partial_\star^a\triangleright(f\,g)\neq
f\,(\partial_\star^a\triangleright g)+(\partial_\star^a\triangleright
f)\,g$ in general, but rather the generalized Leibniz rules spelled
out in the previous section in order to achieve consistency for
$\theta\neq0$. Let us now construct the three sets of derivatives of
interest to us here.

\subsection{Time Ordering \label{TOderiv}}

For the time ordered case, we use (\ref{TOstargen}) to compute the
star-products
\bea
x^-*f&=&\left(x^--\ii\theta\,\mz\cdot\mdell+\ii\theta\,
\overline{\mz}\cdot\overline{\mdell}\,\right)\,f \ , \nonumber\\
x^+*f&=&x^+\,f \ , \nonumber\\
z_i*f&=&\left(z_i-\ii\theta\,x^+\,\overline{\partial}{}^{\,i}
\right)\, f \ , \nonumber\\
\overline{z}_i*f&=&\left(\,\overline{z}_i+\ii\theta\,x^+\,\partial^i
\right)\,f \ .
\label{TOxfprods}\eea
Substituting these into (\ref{dxreq}) using (\ref{eq:rho:nw4}) then
shows that the actions of the $*$-derivatives simply coincide with the
canonical actions of the translation generators on
$\CC^\infty(\mfn^{\vee\,})$, so that
\beq
\partial_*^a\triangleright f=\partial^af \ .
\label{TOderivs}\eeq
Thus the time-ordered noncommutative geometry of $\NW_6$ is invariant
under {\it ordinary} translations of the spacetime in all coordinate
directions, with the generators obeying the twisted Leibniz
rules~(\ref{TOLeibniz}).

\subsection{Symmetric Time Ordering \label{STOderiv}}

Next, consider the case of symmetric time ordering. From
(\ref{TOsymstargen}) we compute the star-products
\bea
x^-\bullet f&=&\left(x^--\mbox{$\frac{\ii\theta}2$}\,
\mz\cdot\mdell+\mbox{$\frac{\ii\theta}2$}\,\overline{\mz}
\cdot\overline{\mdell}\,\right)\,f \ , \nonumber\\
x^+\bullet f&=&x^+\,f \ , \nonumber\\
z_i\bullet f&=&\e^{\frac{\ii\theta}2\,\partial_-}\,
\left(z_i-\ii\theta\,x^+\,\overline{\partial}{}^{\,i}
\right)\,f  \ , \nonumber\\
\overline{z}_i\bullet f&=&\e^{-\frac{\ii\theta}2\,\partial_-}\,
\left(\,\overline{z}_i+\ii\theta\,
x^+\,\partial^i\right)\,f \ .
\label{STOxfprods}\eea
Substituting (\ref{STOxfprods}) into (\ref{dxreq}) using
(\ref{eq:rho:nw4}) along with the derivative rule
\beq
\e^{\ii\theta\,\partial_-}x^-=(x^-+\ii\theta)~\e^{\ii\theta\,
\partial_-} \ ,
\label{derivrule1}\eeq
we find that the actions of the $\bullet$-derivatives on
$\CC^\infty(\mfn^{\vee\,})$ are generically non-trivial and given by
\bea
\partial_-^\bullet\triangleright f&=&\partial_-f \ , \nonumber\\
\partial_+^\bullet\triangleright f&=&\partial_+f \ , \nonumber\\
\partial^i_\bullet\triangleright f&=&\e^{-\frac{\ii\theta}2\,
\partial_-}\,\partial^if \ , \nonumber\\
\overline{\partial}{}^{\,i}_\bullet\triangleright f&=&
\e^{\frac{\ii\theta}2\,\partial_-}\,
\overline{\partial}{}^{\,i}f \ .
\label{STOderivs}\eea
Only the transverse space derivatives are modified owing to the fact
that the Brinkman coordinate system is invariant under translations of
the light-cone coordinates $x^\pm$. Again the twisted Leibniz rules
(\ref{STOLeibniz}) are straightforward to verify in this instance.

\subsection{Weyl Ordering \label{Weylderiv}}

Finally, from the Weyl-ordered star-product (\ref{Weylstargen}) we
compute
\bea
x^-\star f&=&\left[x^-+\left(1-\frac1{\phi_{-\theta}(\ii\partial_-)}
\right)\,\frac{\mz\cdot\mdell}{\partial_-}+
\left(1-\frac1{\phi_{\theta}(\ii\partial_-)}
\right)\,\frac{\overline{\mz}\cdot\overline{\mdell}}{\partial_-}
\right.\nonumber\\ &&\qquad-\left.2\theta\,x^+\,\left(\frac{2}
{\theta\,\partial_-}-\cot\left(\mbox{$\frac\theta2\,\partial_-$}
\right)\right)\,\frac{\overline{\mdell}
\cdot\mdell}{\partial_-}\right]\,f \ , \nonumber\\
x^+\star f&=&x^+\,f \ , \nonumber\\
z_i\star f&=&\left[\frac{z_i}{\phi_{-\theta}(\ii\partial_-)}+
2x^+\,\left(1-\frac1{\phi_{-\theta}(\ii\partial_-)}\right)\,
\frac{\overline{\partial}{}^{\,i}}{\partial_-}\right]\,f \ ,
\nonumber\\ \overline{z}_i\star f&=&
\left[\frac{\overline{z}_i}{\phi_{\theta}(\ii\partial_-)}+
2x^+\,\left(1-\frac1{\phi_{\theta}(\ii\partial_-)}\right)\,
\frac{\partial^i}{\partial_-}\right]\,f \ .
\label{Weylxfprods}\eea
{}From (\ref{dxreq}), (\ref{eq:rho:nw4}) and the derivative rule
\beq
\phi_\theta(\ii\partial_-)x^-=
\frac{\e^{\ii\theta\,\partial_-}-\phi_\theta(\ii\partial_-)}
{\ii\partial_-}+x^-\,\phi_\theta(\ii\partial_-) \ ,
\label{derivrule2}\eeq
it then follows that the actions of the $\star$-derivatives on
$\CC^\infty(\mfn^{\vee\,})$ are given by
\bea
\partial_-^\star\triangleright f&=&\partial_-f \ , \nonumber\\
\partial_+^\star\triangleright f&=&\left[\partial_++2\,
\left(1-\frac{\sin(\theta\,\partial_-)}{\theta\,\partial_-}
\right)\,\frac{\overline{\mdell}\cdot\mdell}{\partial_-}
\right]f \ , \nonumber\\ \partial_\star^i\triangleright f&=&
-\frac{1-\e^{\ii\theta\,\partial_-}}{\ii\theta\,\partial_-}\,
\partial^if \ , \nonumber\\ \overline{\partial}{}_\star^{\,i}
\triangleright f&=&\frac{1-\e^{-\ii\theta\,\partial_-}}
{\ii\theta\,\partial_-}\,\overline{\partial}{}^{\,i}f \ .
\label{Weylderivs}\eea
Thus in the completely symmetric noncommutative geometry of $\NW_6$ both the
light-cone and the transverse space of the plane wave are generically
only invariant under rather complicated twisted translations, obeying
the involved Leibniz rules (\ref{WeylLeibniz}).

\newsection{Traces\label{Integrals}}

The final ingredient required to construct noncommutative field theory
action functionals is a definition of integration. At the algebraic
level, we define an integral to be a
trace on the algebra $\overline{U(\mfn)^\complex}$, i.e. a map
$\ncint:\overline{U(\mfn)^\complex}\to\complex$ which is linear,
\beq
\ncint\bigl(c_1\,\Omega(f)+c_2\,\Omega(g)\bigr)=
c_1\,\ncint\Omega(f)+c_2\,\ncint\Omega(g)
\label{ncintlin}\eeq
for all $f,g\in\CC^\infty(\mfn^{\vee\,})$ and $c_1,c_2\in\complex$, and which
is cyclic,
\beq
\ncint\Omega(f)\cdot\Omega(g)=\ncint\Omega(g)\cdot\Omega(f) \ .
\label{ncintcyclic}\eeq
We define the integral in the star-product formalism using the usual
definitions for the integration of commuting Schwartz functions in
$\CC^\infty(\R^6)$. Then the linearity property (\ref{ncintlin}) is
automatically satisfied. To satisfy the cyclicity requirement
(\ref{ncintcyclic}), we
introduce~\cite{CalWohl1,BehrSyk1,AA-CAA1,DJMTWW1,FelShoi1} a measure
$\kappa$ on $\R^6$ which deforms the flat space volume element
$\dd\mbf x$ and define
\beq
\ncint\Omega(f):=\int\limits_{\R^6}\,\dd\mbf x~\kappa(\mbf x)~f(\mbf x) \ .
\label{ncintdef}\eeq
The measure $\kappa$ is chosen in order to achieve the property
(\ref{ncintcyclic}), so that
\beq
\int\limits_{\R^6}\,\dd\mbf x~\kappa(\mbf x)~(f\star g)(\mbf x)=
\int\limits_{\R^6}\,\dd\mbf x~\kappa(\mbf x)~(g\star f)(\mbf x) \ .
\label{mucyclic}\eeq
Such a measure always exists~\cite{CalWohl1,DJMTWW1,FelShoi1} and its
inclusion in the present context is natural for the curved spacetime
$\NW_6$ which we are considering here. It is important note that, for
the star-products that we use, a measure which satisfies
(\ref{mucyclic}) gives the integral the additional property
\beq
\int\limits_{\R^6}\,\dd\mbf x~\kappa(\mbf x)~(f\star g)(\mbf x)=
\int\limits_{\R^6}\,\dd\mbf x~\kappa(\mbf x)~f(\mbf x)\,g(\mbf x) \ ,
\label{ncintaddprop}\eeq
providing an explicit realization of the Connes-Flato-Sternheimer
conjecture~\cite{FelShoi1}.

Since the coordinate functions $x_a$ generate the
noncommutative algebra, the cyclicity constraint (\ref{mucyclic}) is
equivalent to the star-commutator condition
\beq
\int\limits_{\R^6}\,\dd\mbf x~\kappa(\mbf x)~\bigl[(x_a)^n\,,\,f(\mbf x)
\bigr]_\star=0
\label{starcommcond}\eeq
which must hold for arbitrary functions $f\in\CC^\infty(\R^6)$ (for
which the integral makes sense) and for
all $n\in\N$, $a=1,\dots,6$. Expanding the star-commutator bracket
using its derivation property brings (\ref{starcommcond}) to the form
\beq
\int\limits_{\R^6}\,\dd\mbf x~\kappa(\mbf x)~\sum_{m=0}^n\,{n\choose
  m}\,(x_a)^{n-m}\star\bigl[x_a\,,\,f(\mbf x)\bigr]_\star\star
(x_a)^m=0 \ .
\label{commcondexp}\eeq
We may thus insert the explicit form of $[x_a,f]_\star$ for generic
$f$ and use the ordinary integration by parts property
\beq
\int\limits_{\R^6}\,\dd\mbf x~f(\mbf x)\,g(\mbf x)\,(\partial^a)^nh(\mbf x)=
(-1)^n\,\int\limits_{\R^6}\,\dd\mbf x~\bigl(f(\mbf x)\,(\partial^a)^n
g(\mbf x)\,
h(\mbf x)+(\partial^a)^nf(\mbf x)\,g(\mbf x)\,h(\mbf x)\bigr)
\label{intpartsfgh}\eeq
for Schwartz functions $f,g,h\in\CC^\infty(\R^6)$. This will lead to a
number of constraints on the measure~$\kappa$.

The trace (\ref{ncintdef}) can also be used to define an inner product
$(-,-):\CC^\infty(\mfn^{\vee\,})\times\CC^\infty(\mfn^{\vee\,})\to\complex$
through
\beq
(f,g):=\int\limits_{\R^6}\,\dd\mbf x~\kappa(\mbf x)~\bigl(\,\overline{f}\star
g\bigr)(\mbf x) \ .
\label{ncintinnprod}\eeq
Note that this is different from the inner product introduced in
Section~\ref{Defs}. When we come to deal with the variational
principle in the next section, we shall require that our
star-derivative operators $\partial^a_\star$ be anti-hermitean with
respect to the inner product (\ref{ncintinnprod}),
i.e. $(f,\partial^a_\star\triangleright
g)=-(\partial^a_\star\triangleright f,g)$, or equivalently
\beq
\int\limits_{\R^6}\,\dd\mbf x~\kappa(\mbf x)~\bigl(\,\overline{f}\star
\partial_\star^a
\triangleright g\bigr)(\mbf x)=-\int\limits_{\R^6}\,\dd\mbf x~\kappa(\mbf x)~
\bigl(\,\overline{\partial_\star^a\triangleright f}\star g\bigr)(\mbf x) \ .
\label{ncintparts}\eeq
This allows for a generalized integration by parts
property~\cite{DJMTWW1} for our noncommutative integral. As always, we
will now go through our list of star-products to explore the
properties of the integral in each case. We will find that the measure
$\kappa$ is not uniquely determined by the above criteria and there is
a large flexibility in the choices that can be made. We will also find
that the derivatives of the previous section must be generically
modified by a $\kappa$-dependent shift in order to
satisfy~(\ref{ncintparts}).

\subsection{Time Ordering\label{TOint}}

Using (\ref{TOxfprods}) along with the analogous $*$-products $f*x_a$
we arrive at the $*$-commutators
\begin{eqnarray}
 \cb{x^-}{f}_\ast&=&\ii\theta\,\left(\,\obfz\cdot\obfd
    -\bfz\cdot\bfd\right)f \ , \nn \\ \nn
  \cb{x^+}{f}_\ast&=&0 \ , \\ \nn
  \cb{z_i}{f}_\ast&=&z_i\,\left(1
    -\e^{-\ii\theta\,\d_-}\right)f-\ii\theta\,x^+\,\left(1
    +\e^{-\ii\theta\,\d_-}\right)\od{}^{\,i}f \ , \\
  \cb{\,\oz_i}{f}_\ast&=&\oz_i\,\left(1
    -\e^{\ii\theta\,\d_-}\right)f+\ii\theta\,x^+\,\left(1
    +\e^{\ii\theta\,\d_-}\right)\d^if \ .
\label{eq:time:comm}\end{eqnarray}
When inserted into \eqref{commcondexp}, after integration by parts and
application of the derivative rule (\ref{derivrule1})
these expressions imply constraints on the corresponding measure
$\kappa_\ast$ given by
\begin{eqnarray}
\left(1-\e^{\ii\theta\,\d_-}\right)\kappa_\ast&=&0 \ , \nn \\
\left(1+\e^{\ii\theta\,\d_-}\right)\od{}^{\,i}\kappa_\ast&=&0 \ , \nn \\ \nn
\left(1-\e^{-\ii\theta\,\d_-}\right)\d^i\kappa_\ast&=&0 \ , \\
  \bfz\cdot\bfd\kappa_\ast&=&\obfz\cdot\obfd\kappa_\ast \ .
 \label{eq:time:mu:all}\end{eqnarray}
It is straightforward to see that the equations (\ref{eq:time:mu:all})
imply that the measure must be independent of both the light-cone
position and transverse coordinates, so that
\begin{equation}
  \label{eq:time:mu}
  \partial_-\kappa_\ast=\d^i\kappa_\ast=\od{}^{\,i}\kappa_\ast=0 \ .
\end{equation}

However, the derivative $\d_+^\ast$ in \eqref{TOderivs} does not
satisfy the anti-hermiticity requirement \eqref{ncintparts}. This can be
remedied by translating it by a logarithmic derivative of the measure
$\kappa_*$ and defining the modified $*$-derivative
\beq
  \label{eq:time:d}
 \widetilde\d{}^{\,\ast}_+=\d_+ + \mbox{$\frac12$}\,\d_+\ln\kappa_\ast \ .
\eeq
The remaining $*$-derivatives in (\ref{TOderivs}) are unaltered. While
this redefinition has no adverse effects on the commutation relations
\eqref{eq:rho:nw4}, the action
$\widetilde\d{}^{\,\ast}_+\triangleright f$ contains an additional
linear term in $f$ even if the function $f$ is independent of the time
coordinate $x^+$.

\subsection{Symmetric Time Ordering\label{STOint}}

Using \eqref{STOxfprods} along with the corresponding
$\bullet$-products $f\bullet x_a$ we arrive at the
$\bullet$-commutators
\begin{eqnarray}
\nn  \cb{x^-}{f}_\bullet&=&\ii\theta\,\left(\,
\obfz\cdot\obfd -\bfz\cdot\bfd\right)f \ , \\ \nn
  \cb{x^+}{f}_\bullet&=&0 \ , \\ \nn
  \cb{z_i}{f}_\bullet&=&2\ii
  z_i\,\sin\left(\mbox{$\frac\theta2$}\,\partial_-
\right)f-2\ii\theta\,x^+\,\od{}^{\,i}
\cos\left(\mbox{$\frac\theta2$}\,\partial_-\right)f \ , \\
  \cb{\,\oz_i}{f}_\bullet&=&-2\ii\oz_i\,\sin\left(\mbox{$\frac\theta2$}
\,\partial_-\right)f+2\ii\theta\,x^+\d^i\cos
\left(\mbox{$\frac\theta2$}\,\partial_-\right)f \ .
  \label{eq:symtime:comm}\end{eqnarray}
Substituting these into \eqref{commcondexp} and integrating by parts,
we arrive at constraints on the measure $\kappa_\bullet$ given by
\begin{eqnarray}
 \nn  \bigl(1-\od{}^{\,i}\bigr)\sin\left(\mbox{$\frac\theta2$}\,
\partial_-\right)\kappa_\bullet&=&0 \ , \\ \nn
  \bigl(1+\d^i\bigr)\sin\left(\mbox{$\frac\theta2$}\,
\partial_-\right)\kappa_\bullet&=&0 \ , \\
  \bfz\cdot\bfd\kappa_\bullet&=&\obfz\cdot\obfd\kappa_\bullet
 \label{eq:symtime:mu:all}\end{eqnarray}
which can be reduced to the conditions
\begin{equation}
  \label{eq:symtime:mu:rest}
  \bfz\cdot\bfd\kappa_\bullet=\obfz\cdot\obfd\kappa_\bullet \ ,
  \quad \d_-\kappa_\bullet=0 \ .
\end{equation}
Now the derivative operators $\d_+^\bullet$, $\d^i_\bullet$ and
$\od{}^{\,i}_\bullet$ all violate the requirement
\eqref{ncintparts}. Introducing translates of $\d^i_\bullet$ and
$\od{}^{\,i}_\bullet$ analogously to what we did in (\ref{eq:time:d})
is problematic. While such a shift does not alter the canonical
commutation relations between the coordinates and derivatives,
i.e. the algebraic properties of the differential operators, it does
violate the $\bullet$-commutator relationships \eqref{dxreq} and
\eqref{eq:rho:nw4} for generic functions $f$. Consistency between
differential operator and function commutators would only be
possible in this case by demanding that multiplication from the left
follow a Leibniz-like rule for the translated part.

Thus in order to satisfy both sets of constraints, we are forced to
further require that the measure $\kappa_\bullet$ depend only on the
plane wave time coordinate $x^+$ so that (\ref{eq:symtime:mu:rest})
truncates to
\begin{equation}
  \label{eq:symtime:mu}
  \d^i\kappa_\bullet=\od{}^{\,i}\kappa_\bullet=\d_-\kappa_\bullet=0 \ .
\end{equation}
The logarithmic translation of $\d_+^\bullet$ must still be applied in
order to ensure that the time derivative is anti-hermitean with
respect to the noncommutative inner product. This modifies its action
to
\beq
  \label{eq:symtime:d}
  \widetilde\d{}^{\,\bullet}_+=\d_++\mbox{$\frac12$}\,\d_+\ln\kappa_\bullet
\ .
\eeq
 The actions of all other $\bullet$-derivatives are as in
 \eqref{STOderivs}. Again this shifting has no adverse effects on
 \eqref{eq:rho:nw4}, but it carries the same warning as in the time
 ordered case regarding extra linear terms from the action
 $\widetilde\d{}^{\,\bullet}_+\triangleright f$.

\subsection{Weyl Ordering\label{Weylint}}

Finally, the Weyl ordered star-products \eqref{Weylxfprods} along with
the corresponding $f\star x_a$ products lead to the
$\star$-commutators
\begin{eqnarray}
  \nn \cb{x^-}{f}_\star&=&\ii\theta\,\left(\,\obfz\cdot\obfd
    -\bfz\cdot\bfd\right)f \ , \\ \nn
  \cb{x^+}{f}_\star&=&0 \ , \\ \nn
   \cb{z_i}{f}_\star&=&\ii\theta\,\left(z_i\,\d_-
    -2x^+\,\od{}^{\,i}\right)f \ , \\
  \cb{\,\oz_i}{f}_\star&=&\ii\theta\,\left(-\oz_i\,\d_-
    +2x^+\,\d^i\right)f \ .
\label{eq:weyl:comm}\end{eqnarray}
Substituting these commutation relations into \eqref{commcondexp},
integrating by parts, and using the derivative rules
(\ref{derivrule1}) and (\ref{derivrule2}) leads to the corresponding
measure constraints
\begin{eqnarray}
   \nn  z_i\,\d_-\kappa_\star&=&2x^+\,\od{}^{\,i}\kappa_\star \ , \\ \nn
  \oz_i\,\d_-\kappa_\star&=&2x^+\,\d^i\kappa_\star \ , \\
  \bfz\cdot\bfd\kappa_\star&=&\obfz\cdot\obfd\kappa_\star \ .
\label{eq:weyl:mu:all}\end{eqnarray}
Again these differential equations imply that the measure
$\kappa_\star$ depends only on the plane wave time coordinate $x^+$ so
that
\begin{equation}
  \label{eq:weyl:mu}
  \d_-\kappa_\star=\d^i\kappa_\star=\od{}^{\,i}\kappa_\star=0 \ .
\end{equation}
Translating the derivative operator $\d_+^\star$ as before in order to
satisfy \eqref{ncintparts} yields the modified derivative
\beq
  \label{eq:weyl:d}
  \widetilde\d{}^{\,\star}_+=\d_++2\,
\left(1-\frac{\sin(\theta\,\partial_-)}{\theta\,\partial_-}
\right)\,\frac{\overline{\mdell}\cdot\mdell}{\partial_-}+
\mbox{$\frac12$}\,\d_+\ln\kappa_\star \ ,
\eeq
with the remaining $\star$-derivatives in \eqref{Weylderivs}
unchanged. Once again this produces no major alteration to
\eqref{eq:rho:nw4} but does yield extra linear terms in the actions
$\widetilde\d{}^{\,\star}_+\triangleright f$.

\newsection{Field Theory on $\mbf{\NW_6}$\label{FieldTheory}}

We are now ready to apply the detailed constructions of the preceding
sections to the analysis of noncommutative field theories on the plane
wave $\NW_6$, regarded as the worldvolume of a non-symmetric
D5-brane~\cite{KNSanjay1}. In this paper we will only study the
simplest example of free scalar fields, leaving the detailed analysis
of interacting field theories and higher spin (fermionic and gauge)
fields for future work. The analysis of this section will set the
stage for more detailed studies of noncommutative field theories in
these settings, and will illustrate some of the generic features that
one can expect.

Given a real scalar field $\varphi\in\CC^\infty(\mfn^{\vee\,})$ of mass $m$,
we define an action functional using the integral (\ref{ncintdef}) by
\beq
S[\varphi]=\int\limits_{\R^6}\,\dd\mbf x~\kappa(\mbf x)~\left[
\mbox{$\frac12$}\,\eta_{ab}\,\bigl(\,\widetilde
{\partial}{}^{\,a}_{\star}\triangleright\varphi\bigr)\star
\bigl(\,\widetilde{\partial}{}^{\,b}_{\star}\triangleright\varphi\bigr)+
\mbox{$\frac12$}\,m^2\,\varphi\star\varphi\right] \ ,
\label{Svarphidef}\eeq
where $\eta_{ab}$ is the invariant Minkowski metric tensor induced by
the inner product (\ref{NW4innerprod}) with the non-vanishing components
$\eta_{\pm\,\mp}=1$ and $\eta_{z_i\,\overline{z}_j}=\frac12\,\delta_{ij}$. The
tildes on the derivatives in (\ref{Svarphidef}) indicate that the time
component must be appropriately shifted as described in the previous
section. Using the property (\ref{ncintaddprop}) we may simplify the
action to the form
\beq
S[\varphi]=\int\limits_{\R^6}\,\dd\mbf x~\kappa(\mbf x)~\left[
\mbox{$\frac12$}\,\eta_{ab}\,\bigl(\,\widetilde
{\partial}{}^{\,a}_{\star}\triangleright\varphi\bigr)
\bigl(\,\widetilde{\partial}{}^{\,b}_{\star}\triangleright\varphi\bigr)+
\mbox{$\frac12$}\,m^2\,\varphi^2\right] \ .
\label{Svarphisimpl}\eeq

By using the integration by parts property (\ref{ncintparts}) on
Schwartz fields $\varphi$, we may easily compute the first order
variation of the action (\ref{Svarphisimpl}) to be
\beq
\frac{\delta S[\varphi]}{\delta\varphi}~\delta\varphi:=
S[\varphi+\delta\varphi]-S[\varphi]=-\int\limits_{\R^6}\,\dd\mbf x~
\kappa(\mbf x)~\left[\eta_{ab}\,\overline{\widetilde
{\partial}{}_\star^{\,a}}\triangleright\bigl(\,\widetilde
{\partial}{}_\star^{\,b}\triangleright\varphi\bigr)-m^2\,\varphi^2
\right]~\delta\varphi \ .
\label{actionvary1}\eeq
Applying the variational principle $\frac{\delta
  S[\varphi]}{\delta\varphi}=0$ to (\ref{actionvary1}) thereby leads
to the noncommutative Klein-Gordan field equation
\beq
\Box^\star\triangleright\varphi-m^2\,\varphi=0
\label{NCeom}\eeq
where
\beq
\Box^\star\triangleright\varphi:=2\,\partial_+\triangleright
\partial_-\varphi+\mdell^\top\triangleright
\overline{\mdell}\triangleright\varphi+\mbox{$\frac12$}\,
\partial_+\ln\kappa~\partial_-\varphi
\label{Boxstardef}\eeq
and we have used $\partial_-\kappa=0$. The second order
$\star$-differential operator $\Box^\star$ should be regarded as a deformation
of the covariant Laplace operator $\Box_0^\star$ corresponding to the
commutative plane wave geometry of $\NW_6$. This Laplacian coincides
with the quadratic Casimir element
\beq
{\sf C}:=\theta^{-2}\,\eta^{ab}\,\X_a\,\X_b
=2\,\J\,\T+\mbox{$\frac12\,\sum\limits_{i=1,2}$}\,\bigl(
\P_+^i\,\P_-^i+\P_-^i\,\P_+^i\bigr)
\label{quadCasimir}\eeq
of the universal enveloping algebra $U(\mfn)$, expressed in terms of
left or right isometry generators for the action of the isometry group
$\mathcal{N}_{\rm L}\times\mathcal{N}_{\rm R}$ on
$\NW_6$~\cite{PK1,CFS1,HSz1}.

However, in the manner which we have constructed things, this is not
the case. Recall that the approximation in which our quantization of
the geometry of $\NW_6$ holds is the small time limit $x^+\to0$ in
which the plane wave approaches flat six-dimensional Minkowski space
$\eucl^{1,5}$. To incorporate the effects of the curved geometry of
$\NW_6$ into our formalism, we have to replace the derivative
operators $\widetilde{\partial}{}_\star^{\,a}$ appearing in
(\ref{Svarphidef}) with appropriate curved space analogs
$\delta_\star^a$~\cite{BehrSyk1,HoMiao1}.

Recall that the derivative operators $\partial_\star^{a}$ are {\it
  not} derivations of the star-product $\star$, but instead obey the
deformed Leibniz rules (\ref{defLeibniz}). The deformation arose from
twisting the co-action of the bialgebra $U(\mfg)$ so that it generated
automorphisms of the noncommutative algebra of functions,
i.e. isometries of the noncommutative plane wave. The basic idea is to
now ``absorb'' these twistings into derivations $\delta_\star^a$
obeying the usual Leibniz rule
\beq
\delta_\star^a\triangleright(f\star g)=\left(\delta_\star^a
\triangleright f\right)\star g+f\star\left(\delta_\star^a
\triangleright g\right) \ .
\label{deltaLeibniz}\eeq
These derivations generically act on $\CC^\infty(\mfn^{\vee\,})$ as the
noncommutative $\star$-polydifferential operators
\beq
\delta_\star^a\triangleright f=\sum_{n=1}^\infty\,\xi_a{}^{a_1\cdots
  a_n}\star\left(\partial_{a_1}^\star\triangleright\cdots\triangleright
\partial_{a_n}^\star\triangleright f\right)
\label{polydiffops}\eeq
with $\xi_a{}^{a_1\cdots a_n}\in\CC^\infty(\mfn^{\vee\,})$. Unlike the
derivatives $\partial_\star^a$, these derivations will no longer
star-commute among each other. There is a one-to-one
correspondence~\cite{Kont1} between such derivations $\delta_a^\star$
and Poisson vector fields $E^a=E^a{}_b~\partial^b$ on $\mfn^{\vee\,}$
obeying
\beq
E^a\circ\Theta(f,g)=\Theta(E^af,g)+\Theta(f,E^ag)
\label{Poissonvecfields}\eeq
for all $f,g\in\CC^\infty(\mfn^{\vee\,})$. To leading order one has
$\delta_\star^a\triangleright
f=E^a{}_b\star(\partial_\star^b\triangleright f)+O(\theta)$. By
identifying the Lie algebra $\mfn$ with the tangent space to $\NW_6$,
at this order the vector fields $E^a$ can be thought of as defining a
natural local frame with flat metric $\eta_{ab}$ and a curved metric tensor
$G^\star_{ab}=\frac12\,\eta_{cd}\,(E^c{}_a\star E^d{}_b+E^d{}_a\star
E^c{}_b)$ on the noncommutative space $\NW_6$. However, for our
star-products there are always higher order terms in
(\ref{polydiffops}) which spoil this interpretation. The
noncommutative frame fields $\delta_\star^a$ describe the {\it
  quantum} geometry of the plane wave $\NW_6$. In particular, the
metric tensor $G^\star$ will in general differ from the
classical open string metric $G_{\rm open}$. While the operators
$\delta_\star^a$ always exist as a consequence of the Kontsevich
formality map~\cite{Kont1,BehrSyk1}, computing them explicitly is a
highly difficult problem. We will see some explicit examples below, as
we now begin to tour through our three star-products. Throughout we
shall take the natural choice of measure $\kappa=\sqrt{|\det
  G|}=\frac12$, the constant Riemannian volume density of the $\NW_6$
plane wave geometry.

\subsection{Time Ordering\label{ScalarTO}}

In the case of time ordering, we use (\ref{TOderivs}) to compute
\beq
\Box^*\triangleright\varphi=\left(2\,\partial_+\,\partial_-+
\overline{\mdell}\cdot\mdell\right)\varphi
\label{TOBoxeq}\eeq
and thus the equation of motion coincides with that of a free scalar
particle on flat Minkowski space $\eucl^{1,5}$ (Deviations from
flat spacetime can only come about here by choosing a time-dependent measure
$\kappa_*$). This illustrates the point made above that the treatment
of the present paper tackles only the semi-classical flat space limit of the
spacetime $\NW_6$. The appropriate curved geometry for this ordering
corresponds to the global coordinate system (\ref{NW4metricNW}) in
which the classical Laplace operator is given by
\beq
\Box_0^*=2\,\partial_+\,\partial_-+
\left|\mdell+\mbox{$\frac\ii2$}\,\theta\,\overline{\mz}\,
\partial_-\right|^2 \ ,
\label{TOBox0}\eeq
so that the free wave equation $(\Box_0^*-m^2)\varphi=0$ is equivalent
to the Schr\"odinger equation for a particle of charge $p^+$ (the
momentum along the $x^-$ direction) in a constant magnetic field
of strength~$\theta$. A global pseudo-orthonormal frame is provided by
the commutative vector fields
\bea
E_-^*&=&\partial_- \ , \nn\\ E_+^*&=&\partial_+-\ii\theta\,\left(
\mz\cdot\mdell-\overline{\mz}\cdot\overline{\mdell}\,\right) \ ,
\nn\\ E^{i}_*&=&\partial^i \ , \nn\\
\overline{E}{}_{*}^{\,i}&=&\overline{\partial}{}^{\,i} \ .
\label{TOorthoframe}\eea

Determining the derivations $\delta_*^a$ corresponding to the
commuting frame (\ref{TOorthoframe}) on the quantum space is in
general rather difficult. Evidently, from the coproduct structure
(\ref{TOLeibniz}) the action along the light-cone position is given by
\beq
\delta_-^*\triangleright f=\partial_-f \ .
\label{TOdeltaminus}\eeq
This is simply a consequence of the fact that translations along $x^-$
generate an automorphism of the noncommutative algebra of functions,
i.e. an isometry of the noncommutative geometry. From the Hopf algebra
coproduct (\ref{TOcoprods}) we have
\beq
\Delta_*\bigl(\e^{\ii\theta\,\partial_-}\bigr)=
\e^{\ii\theta\,\partial_-}\otimes\e^{\ii\theta\,\partial_-}
\label{TOcoprodglobal}\eeq
and consequently
\beq
\e^{\ii\theta\,\partial_-}\triangleright(f*g)=\bigl(
\e^{\ii\theta\,\partial_-}\triangleright f\bigr)*\bigl(
\e^{\ii\theta\,\partial_-}\triangleright g\bigr) \ .
\label{xmautoNC}\eeq

On the other hand, the remaining isometries involve intricate
twistings between the light-cone and transverse space directions. For
example, let us demonstrate how to unravel the coproduct rule for
$\partial_+^*$ in (\ref{TOLeibniz}) into the desired symmetric Leibniz
rule (\ref{deltaLeibniz}) for $\delta_+^*$. This can be achieved by
exploiting the $*$-product identities
\bea
z_i*f&=&\bigl(\e^{\ii\theta\,\partial_-}f\bigr)*z_i-2\ii\theta\,
x^+\,\overline{\partial}{}^{\,i}f \ , \nn\\
\overline{z}_i*f&=&\bigl(\e^{-\ii\theta\,\partial_-}f\bigr)*
\overline{z}_i+2\ii\theta\,x^+\,\partial^if
\label{TOstarprodcomm}\eea
along with the commutativity properties
$[\partial_-^*,z_i]_*=[\partial_-^*,\overline{z}_i]_*=0$ for $i=1,2$
and for arbitrary functions $f$. Using in addition the modified
Leibniz rules (\ref{TOLeibniz}) along with the $*$-multiplication
properties (\ref{TOxfprods}) we thereby find
\beq
\delta_+\triangleright f=\left[x^+\,\partial_++\mbox{$\frac1{2\ii}$}\,
\left(\mz\cdot\overline{\mdell}+\overline{\mz}\cdot\mdell\right)
\right]f \ .
\label{TOdeltap}\eeq
This action mimicks the form of the classical frame field $E_+^*$ in
(\ref{TOorthoframe}).

Finally, for the transversal isometries, one can attempt to seek
functions $g^i\in\CC^\infty(\mfn^{\vee\,})$ such that
$g^i*f=(\e^{-\ii\theta\,\partial_-}f)*g_i$ in order to absorb
the light-cone translation in the Leibniz rule for $\partial_*^i$ in
(\ref{TOLeibniz}). This would mean that the $x^-$ translations are
generated by {\it inner} automorphisms of the noncommutative
algebra. If such functions exist, then the corresponding derivations
are given by $\delta^i_*\triangleright f=g^i*\partial_*^if$ (no sum over
$i$) and similarly for $\overline{\delta}{}^{\,i}_*$. However, it is
doubtful that such inner derivations exist and the transverse space
frame fields are more likely to be given by higher-order $*$-polyvector
fields. For example, using similar steps to those which led to
(\ref{TOdeltap}), one can show that the actions
\bea
\delta_*\triangleright f&:=&\left(\,\overline{\mz}\cdot\mdell+2\ii
x^+\,\partial_+-\ii\theta\,x^+\,\overline{\mdell}\cdot\mdell\right)f
\ , \nn\\
\overline{\delta}{}_*\triangleright f&:=&\left({\mz}\cdot
\overline{\mdell}-2\ii
x^+\,\partial_++\ii\theta\,x^+\,\overline{\mdell}\cdot\mdell\right)f
\label{TOdeltatransv}\eea
define derivations of the $*$-product on $\NW_6$, and hence naturally
determine elements of a noncommutative transverse frame.

The action of the corresponding noncommutative Laplacian
$\eta_{ab}\,\delta_*^a\triangleright(\delta_*^b\triangleright\varphi)$
deforms the harmonic oscillator dynamics generated by (\ref{TOBox0})
by non-local higher spatial derivative terms. These extra terms will
have significant ramifications at large energies for motion in the
transverse space. This could have profound physical effects in the
interacting noncommutative quantum field theory. In particular, it may
alter the UV/IR mixing story~\cite{MVRS1} in an interesting way. For
time-dependent noncommutativity with standard tree-level propagators,
UV/IR mixing becomes intertwined with violations of energy
conservation in an intriguing way~\cite{BG1,RS1}, and it would be
interesting to see how our modified free field propagators affect this
analysis. It would also be interesting to see if and how these
modifications are related to the generic connection between wave
propagation on homogeneous plane waves and the Lewis-Riesenfeld theory
of time-dependent harmonic oscillators~\cite{BlauOL1}.

\subsection{Symmetric Time Ordering\label{ScalarSTO}}

The analysis in the case of symmetric time ordering is very similar to
that just performed, so we will be very brief and only highlight the
essential changes. From (\ref{STOderivs}) we find once again that the
Laplacian (\ref{Boxstardef}) concides with the flat space wave
operator
\beq
\Box^\bullet\triangleright\varphi=\left(2\,\partial_+\,\partial_-+
\overline{\mdell}\cdot\mdell\right)\varphi \ .
\label{STOBoxeq}\eeq
The relevant coordinate system in this case is given by the Brinkman
metric (\ref{NW4metricBrink}) for which the classical Laplace
operator reads
\beq
\Box_0^\bullet=2\,\partial_+\,\partial_-+\overline{\mdell}\cdot
\mdell-\mbox{$\frac14$}\,\theta^2\,|\mz|^2\,\partial_-^2 \ .
\label{STOBox0}\eeq
A global pseudo-orthonormal frame in this case is provided by the
vector fields
\bea
E_-^\bullet&=&\partial_- \ , \nn\\ E_+^\bullet&=&\partial_++
\mbox{$\frac18$}\,\theta^2\,|\mz|^2\,\partial_- \ ,
\nn\\ E^{i}_\bullet&=&\partial^i \ , \nn\\
\overline{E}{}_{\bullet}^{\,i}&=&\overline{\partial}{}^{\,i} \ .
\label{STOorthoframe}\eea
The corresponding twisted derivations $\delta^a_\bullet$ which
symmetrize the Leibniz rules (\ref{STOLeibniz}) can be constructed
analogously to those of the time ordering case in
Section~\ref{ScalarTO} above.

\subsection{Weyl Ordering\label{ScalarWeyl}}

Finally, the case of Weyl ordering is particularly interesting because
the effects of curvature are present even in the flat space
limit. Using (\ref{Weylderivs}) we find the Laplacian
\beq
\Box^\star\triangleright\varphi=\left(2\,\partial_+\,\partial_-
+2\,\left[2\,\left(1-\frac{\sin(\theta\,\partial_-)}
{\theta\,\partial_-}\right)+\frac{1-\cos(\theta\,\partial_-)}
{\theta^2\,\partial_-^2}\right]\,\overline{\mdell}\cdot\mdell
\right)\varphi
\label{WeylBoxeq}\eeq
which coincides with the flat space Laplacian only at $\theta=0$. To
second order in the deformation parameter $\theta$, the equation of
motion (\ref{NCeom}) thereby yields a second order correction to the
usual flat space Klein-Gordan equation given by
\beq
\left[\left(2\,\partial_+\,\partial_-+\overline{\mdell}\cdot\mdell-m^2
\right)+\mbox{$\frac7{12}$}\,\theta^2\,\partial_-^2\,\overline{\mdell}\cdot
\mdell+O\left(\theta^4\right)\right]\varphi=0 \ .
\label{KGeqcorr}\eeq

Again we find that only the transverse space motion is
altered by noncommutativity, but this time through a non-local
dependence on the light-cone momentum $p^+$ yielding a drastic
modification of the dispersion relation for free wave propagation in
the noncommutative spacetime. This dependence is
natural. The classical mass-shell condition for motion in the curved
background is $2\,p^+\,p^-+|4\,\theta\,p^+\,\mbf\lambda|^2=m^2$, where
$\mbf\lambda\in\complex^2$ represents the position and radius of the
circular trajectories in the background magnetic
field~\cite{CFS1}. Thus the quantity $4\,\theta\,p^+\,\mbf\lambda$ can be
interpreted as the momentum for motion in the transverse space. The
operator (\ref{WeylBoxeq}) incorporates the appropriate noncommutative
deformation of this motion. It illustrates the point that the
fundamental quanta governing the interactions in the present class of
noncommutative quantum field theories are not likely to involve the
particle-like dipoles of the flat space cases~\cite{Sheikh1,BigSuss1},
but more likely string-like objects owing to the nonvanishing $H$-flux in
(\ref{NS2formBrink}). These open string quanta become polarized as
dipoles experiencing a net force due to their couplings to the
non-uniform $B$-field. It is tempting to speculate that, in contrast
to the other orderings, the Weyl ordering naturally incorporates the
new vacua corresponding to long string configurations which are due
entirely to the time-dependent nature of the background Neveu-Schwarz
field~\cite{BDAKZ1}.

While the Weyl ordered star-product is natural from an algebraic point
of view, it does not correspond to a natural coordinate system for the
plane wave $\NW_6$ due to the complicated form of the group product
rule (\ref{Weylgpprodexpl}) in this case. In particular, the frame
fields in this instance will be quite complicated. Computing the
corresponding twisted derivations $\delta^a_\star$ directly would
again be extremely cumbersome, but luckily we can exploit the
equivalence between the star-products $\star$ and $*$ derived in
Section~\ref{WOP}. Given the derivations $\delta^a_*$ constructed in
Section~\ref{ScalarTO} above, we may use the differential operator
(\ref{Gdiffopexpl}) which implements the equivalence (\ref{WeylTOrel})
to define
\beq
\delta^a_\star\triangleright f:=\mathcal{G}^{~}_\Omega\circ
\delta^a_*\triangleright\bigl(\mathcal{G}_\Omega^{-1}(f)\bigr) \ .
\label{Weyldelta}\eeq
These noncommutative frame fields will lead to the appropriate curved
space extension of the Laplace operator in (\ref{WeylBoxeq}).

\newsection{Worldvolume Field Theories\label{D3Branes}}

In this final section we will describe how to build noncommutative
field theories on regularly embedded worldvolumes of D-branes in the
spacetime $\NW_6$ using the formalism described above. We shall
describe the general technique on a representative example by
comparing the noncommutative field theory on $\NW_6$ which we have
constructed in this paper to that of the noncommutative D3-branes
which was constructed in~\cite{HSz1}. We shall do so in a general
fashion which illustrates how the construction extends to generic
D-branes. This will provide further perspective on the natures of the different
quantizations we have used throughout, and also illustrate the overall
consistency of our results. As we will now demonstrate, we can view
the noncommutative geometry of $\NW_6$, in the manner constructed
above, as a collection of all euclidean noncommutative D3-branes taken
together. This is done by restricting the geometry to obtain the usual
quantization of coadjoint orbits in $\mfn^{\vee\,}$ (as opposed to all of
$\mfn^{\vee\,}$ as described above). This restriction defines an alternative
and more geometrical approach to the quantization of these branes which
does not rely upon working with representations of the Lie group
$\mathcal{N}$, and which is more adapted to the flat space limit
$\theta\to0$. This procedure can be thought of as somewhat opposite to
the philosophy of~\cite{HSz1}, which quantized the geometry of a
non-symmetric D5-brane wrapping $\NW_6$~\cite{KNSanjay1} by viewing it as a
noncommutative foliation by these euclidean D3-branes. Here the
quantization of the spacetime-filling brane in $\NW_6$ has been
carried out independently leading to a much simpler noncommutative
geometry which correctly induces the anticipated worldvolume field
theories on the $\eucl^4$ submanifolds of $\NW_6$.

The euclidean D3-branes of interest wrap the non-degenerate conjugacy
classes of the group $\mathcal{N}$ and are coordinatized by the
transverse space $\mz\in\complex^2\cong\eucl^4$~\cite{SF-OF1}. They
are defined by the spacelike hyperplanes of constant time in $\NW_6$
given by the transversal intersections of the null hypersurfaces
\bea
x^+&=&{\rm constant} \ , \nn\\
x^-+\mbox{$\frac14$}\,\theta\,|\mz|^2\,\cot\left(\mbox{$\frac12$}\,
\theta\,x^+\right)&=&{\rm constant} \ ,
\label{D3subsps}\eeq
independently of the chosen coordinate frame. This describes the brane
worldvolume as a wavefront expanding in a sphere $\Sphere^3$ in the
transverse space. In the semi-classical flat space limit $\theta\to0$,
the second constraint in (\ref{D3subsps}) to leading order becomes
\beq
C:=2\,x^+\,x^-+|\mz|^2={\rm constant} \ .
\label{Cdefconst}\eeq
The function $C$ on $\mfn^{\vee\,}$ corresponds to the Casimir element
(\ref{quadCasimir}) and the constraint (\ref{Cdefconst}) is analogous
to the requirement that Casimir operators act as scalars in irreducible
representations. Similarly, the constraint on the time coordinate
$x^+$ in (\ref{D3subsps}) is analogous to the requirement that the
central element $\T$ act as a scalar operator in any irreducible
representation of $\mathcal{N}$.

Let $\pi:\NW_6\to\eucl^4$ be the projection of the six-dimensional
plane wave onto the worldvolume of the symmetric D3-branes. Let
$\pi^\sharp:\CC^\infty(\eucl^4)\to\CC^\infty(\NW_6)$ be the induced algebra
morphism defined by pull-back $\pi^\sharp(f)=f\circ\pi$. To consistently
reduce the noncommutative geometry from all of $\NW_6$
to its conjugacy classes, we need to ensure that the candidate
star-product on $\mfn^{\vee\,}$ respects the Casimir property of the
functions $x^+$ and $C$, i.e. that $x^+$ and $C$ star-commute with
every function $f\in\CC^\infty(\mfn^{\vee\,})$. Only in that case can the
star-product be consistently restricted from all of $\NW_6$ to a
star-product $\star_{x^+}$ on the conjugacy classes $\eucl^4$ defined
by
\beq
f\star_{x^+}g:=\pi^\sharp(f)\star\pi^\sharp(g) \ .
\label{starxpdef}\eeq
Then one has the compatibility condition
\beq
\iota^\sharp(f\star g)=\iota^\sharp(f)\star_{x^+}\iota^\sharp(g)
\label{compcondWeyl}\eeq
where $\iota^\sharp:\CC^\infty(\NW_6)\to\CC^\infty(\eucl^4)$ is the
pull-back induced by the inclusion map
$\iota:\eucl^4\hookrightarrow\NW_6$. In this case one has an isomorphism
$\CC^\infty(\eucl^4)\cong\CC^\infty(\NW_6)/\mathcal{J}$ of associative
noncommutative algebras~\cite{Waldmann1}, where $\mathcal{J}$ is the
two-sided ideal of $\CC^\infty(\NW_6)$ generated by the Casimir
constraints $(x^+-{\rm constant})$ and $(C-{\rm constant})$. This
procedure is a noncommutative version of Poisson reduction, with the
Poisson ideal $\mathcal{J}$ implementing the geometric requirement
that the Seiberg-Witten bi-vector $\Theta$ be tangent to the conjugacy
classes.

{}From the star-commutators (\ref{eq:time:comm}),
(\ref{eq:symtime:comm}) and (\ref{eq:weyl:comm}) we see that
$[x^+,f]_\star=0$ for all three of our star-products. However, the
condition $[C,f]_\star=0$ is {\it not} satisfied. Although classically
one has the Poisson commutation $\Theta(C,f)=0$, one can only
consistently restrict the star-products by first defining an
appropriate projection of the algebra of functions on $\mfn^{\vee\,}$ onto
the star-subalgebra $\mathcal{C}$ of functions which star-commute with
the Casimir function $C$. One easily computes that $\mathcal{C}$
naturally consists of functions $f$ which are independent of
the light-cone position, i.e. $\partial_-f=0$. Then the projection
$\iota^\sharp$ above may be applied to the subalgebra $\mathcal{C}$
on which it obeys the requisite compatibility condition
(\ref{compcondWeyl}). The general conditions for reduction of
Kontsevich star-products to D-submanifolds of Poisson manifolds are
described in~\cite{CattFel2,CFal1}.

With these projections implicitly understood, one straightforwardly
finds that all three star-products (\ref{TOstargen}),
(\ref{TOsymstargen}) and (\ref{Weylstargen}) restrict to
\beq
f\star_{x^+}g=\mu\circ\exp\left[\ii\theta\,x^+\,\left(
\mdell^\top\otimes\overline{\mdell}-
{\overline{\mdell}}{}^{\,\top}\otimes\mdell\right)\right]f\otimes g
\label{fstargrestrict}\eeq
for functions $f,g\in\CC^\infty(\eucl^4)$. This is just the Moyal
product, with noncommutativity parameter $\theta\,x^+$, on the
noncommutative euclidean D3-branes. It is cohomologically equivalent
to the Voros product which arises from quantizing the conjugacy
classes through endomorphism algebras of irreducible representations
of the twisted Heisenberg algebra $\mfn$, with a normal or Wick
ordering prescription for the generators $\P_\pm^i$~\cite{HSz1}. In
this case, the noncommutative euclidean space arises from a projection
of $U(\mfn)$ in the discrete representation $V^{p^+,p^-}$ whose second
Casimir invariant (\ref{quadCasimir}) is given in terms of light-cone
momenta as ${\sf C}=-2\,p^+\,(p^-+\theta)$ and with $\T=\theta\,p^+$. In this
approach the noncommutativity parameter is naturally the {\it inverse}
of the effective magnetic field $p^+\,\theta$. On the other hand, the
present analysis is a more geometrical approach to the quantization of
symmetric D3-branes in $\NW_6$ which deforms the euclidean worldvolume
geometry by the time parameter $\theta\,x^+$ without resorting to
endomorphism algebras. The relationship between the two sets of
parameters is given by $x^+=p^+\,\tau$, where $\tau$ is the proper
time coordinate for geodesic motion in the pp-wave geometry of
$\NW_6$.

In contrast to the coadjoint orbit quantization~\cite{HSz1}, the
noncommutativity found here matches exactly that predicted from string
theory in the semi-classical limit~\cite{DAK1}, which asserts that the
Seiberg-Witten bi-vector on the D3-branes is given by
$\Theta_{x^+}=\frac\ii2\,\sin(\theta\,x^+)~\mdell^\top\wedge
\overline{\mdell}$. Note that the present analysis also covers as a
special case the degenerate cylindrical null branes located at time
$x^+=0$~\cite{SF-OF1}, for which (\ref{fstargrestrict}) becomes the
ordinary pointwise product $f\star_0g=f\,g$ of worldvolume fields and
as expected these branes support a {\it commutative} worldvolume
geometry. In contrast, the commutative null branes correspond to the
class of continuous representations of the twisted Heisenberg algebra
having quantum number $p^+=0$ which must be dealt with
separately~\cite{HSz1}.

It is elementary to check that the rest of the geometrical constructs
of this paper reduce to the standard ones appropriate for a Moyal
space. By defining
\beq
\partial_{\star_{x^+}}^a\triangleright
f:=\iota^\sharp\circ\partial_\star^a\triangleright\bigl(\pi^\sharp(f)\bigr) \
,
\label{derivxpdef}\eeq
one finds that the actions of the derivatives constructed in
Section~\ref{Derivatives} all reduce to the standard ones of flat
noncommutative euclidean space,
i.e. $\partial_{\star_{x^+}}^i\triangleright f=\partial^if$,
$\overline{\partial}{}^{\,i}_{\star_{x^+}}\triangleright
f=\overline{\partial}{}^{\,i}f$ for
$f\in\CC^\infty(\eucl^4)$. From Section~\ref{Coprod} one recovers the
standard Hopf algebra of these derivatives with trivial coproducts
$\Delta_{\star_{x^+}}$ defined by
\beq
\Delta_{\star_{x^+}}(\nabla_{\star_{x^+}})\triangleright
(f\otimes g):=\bigl(\iota^\sharp\otimes\iota^\sharp\bigr)\circ
\Delta_\star(\nabla_\star)
\triangleright\bigl(\pi^\sharp(f)\otimes\pi^\sharp(g)\bigr) \ ,
\label{Deltaxpdef}\eeq
and hence the symmetric Leibniz rules appropriate to the translational
symmetry of field theory on Moyal space. Consistent
with the restriction to the conjugacy classes, one also has
$\partial_\pm^{\star_{x^+}}\triangleright f=0$.

However, from (\ref{TOcoprodtime}), (\ref{STOcoprodtime}) and
(\ref{Weylcoprodtime}) one finds a non-vanishing co-action of time
translations given by
\beq
\Delta_{\star_{x^+}}\bigl(\partial_+^{\star_{x^+}}\bigr)=
\theta\,\bigl(\mdell_{\star_{x^+}}{}^\top\otimes
\overline{\mdell}_{\star_{x^+}}-
\overline{\mdell}_{\star_{x^+}}{}^\top\otimes\mdell_{\star_{x^+}}\bigr) \ .
\label{Moyalcoprodtime}\eeq
This formula is very natural. The isometries of $\NW_6$
in $\mfg=\mfn_{\rm L}\oplus\mfn_{\rm R}$ corresponding to the number
operator $\J$ of the twisted Heisenberg algebra are generated by the
vector fields~\cite{HSz1} $J_{\rm L}=\theta^{-1}\,\partial_+$ and $J_{\rm
  R}=-\theta^{-1}\,\partial_+-\ii(\mz\cdot\mdell-
\overline{\mz}\cdot\overline{\mdell}\,)=\theta^{-1}\,E_+^*$ (in
Brinkman coordinates). The vector field $J_{\rm L}+J_{\rm R}$
generates rigid rotations in the transverse space. Restricted to the
D3-brane worldvolume, the time translation isometries thus truncate to
rotations of $\eucl^4$ in ${\rm so}(4)$. The coproduct
(\ref{Moyalcoprodtime}) gives the standard twisted co-action of
rotations for the Moyal algebra which define quantum rotational
symmetries of noncommutative euclidean
space~\cite{CPT1,CKNT1,Wess1}. This discussion also drives home the
point made earlier that our derivative operators $\partial_\star^a$
indeed do generate, through their twisted co-actions (Leibniz rules),
quantum isometries of the full noncommutative plane wave.

Finally, a trace on $\CC^\infty(\eucl^4)$ is induced from
(\ref{ncintdef}) by restricting the integral to the submanifold
$\iota:\eucl^4\hookrightarrow\NW_6$ and using the induced measure
$\iota^\sharp(\kappa)$. For the measures constructed in
Section~\ref{Integrals}, $\iota^\sharp(\kappa)$ is always a constant
function on $\eucl^4$ and hence the integration measures all restrict
to the constant volume form of $\eucl^4$.
Thus noncommutative field theories on the spacetime $\NW_6$ consistently
truncate to the anticipated worldvolume field theories on
noncommutative euclidean D3-branes in $\NW_6$, together with the
correct twisted implementation for the action of classical worldvolume
symmetries. The advantage of the present point of view is that many of
the novel features of these canonical Moyal space field theories
naturally originate from the pp-wave noncommutative geometry when the
Moyal space is regarded as a regularly embedded coadjoint orbit in
$\mfn^{\vee\,}$, as described above. Furthermore, the method detailed in this
paper allows a more systematic construction of the deformed
worldvolume field theories of {\it generic} D-branes in $\NW_6$ in the
semi-classical regime, and not just the symmetric branes analysed
here. For instance, the analysis can in principle be applied to
describe the dynamics of symmetry-breaking D-branes which localize
along products of twisted conjugacy classes in the Lie group
$\mathcal{N}$~\cite{Quella1}. However, these branes have yet to be
classified in the case of the gravitational wave $\NW_6$.

\subsection*{Acknowledgments}

We thank J.~Figueroa-O'Farrill, L.~Friedel, J.~Gracia-Bond\'{\i}a,
P.-M.~Ho, G.~Landi, F.~Lizzi, B.~Schroers and S.~Waldmann for helpful
discussions and correspondence. This work was supported in part by the
EU-RTN Network Grant MRTN-CT-2004-005104. The work of S.H. was
supported in part by an EPSRC Postgraduate Studentship. The work of
R.J.S. was supported in part by PPARC Grant PPA/G/S/2002/00478.

\end{document}